\documentclass{article}

\usepackage{amsmath}

\numberwithin{equation}{section}

\usepackage{amsfonts}

\usepackage{slashed}

\usepackage{epsfig}
\usepackage{graphicx}

\def\be{\begin{equation}}
\def\ee{\end{equation}}
\def\bea{\begin{eqnarray}}
\def\eea{\end{eqnarray}}

\def\({\left(}
\def\){\right)}
\def\<{\left<}
\def\>{\right>}

\def\tr{{\mbox{tr}}}
\def\be{\begin{equation}}
\def\ee{\end{equation}}
\def\ben{\begin{eqnarray}}
\def\een{\end{eqnarray}}
\def\({\left(}
\def\){\right)}
\def\<{\left<}
\def\>{\right>}
\def\!{\right|}
\def\|{\left|}

\def\[{\left[}
\def\]{\right]}

\def\x{\right|}
\def\+{\bar}
\def\mb{\mathbb}
\def\tr{{\mbox{tr}}}

\def\L{{\cal{L}}}
\def\t{\widetilde}

\def\M{{\cal{M}}}

\def\N{{\cal{N}}}

\def\E{{\cal{E}}}

\def\M{{\cal{M}}}
\def\Q{{\cal{Q}}}

\def\x{{\vec{x}}}

\def\1{{\dot{1}}}
\def\2{{\dot{2}}}

\begin{document}
\setlength{\unitlength}{1mm}

\pagestyle{empty}
\vskip-10pt
\vskip-10pt
\hfill 
\begin{center}
\vskip 3truecm
{\Large \bf
One dyonic instanton in\\
5d maximal SYM theory}\\
\vskip 2truecm
{\large \bf
Dongsu Bak\footnote{dsbak@uos.ac.kr}, Andreas Gustavsson\footnote{agbrev@gmail.com} }\\
\vskip 1truecm
{\it  Physics Department, University of Seoul,  Seoul 130-743 Korea}
\end{center}
\vskip 2truecm
{\abstract{We study the one-instanton sector of 5d U(N) maximal SYM theory. By using the moduli space approximation we obtain all the 1/4-BPS
bound states of one dyonic instanton when a vev is given to one of the SYM scalar fields that breaks U(N) maximally.
We compute the corresponding 1/4-BPS index and partition function and find agreement with arXiv:1110.2175.}}

\vfill
\vskip4pt
\eject
\pagestyle{plain}

\section{Introduction}

A direct formulation of 6d non-Abelian M5 brane theory has been a long-standing problem. The
recent proposal of \cite{Douglas:2010iu, Lambert:2010iw} that
 M5 brane theory compactified in the M-theory circle direction with radius $R$, is dual to 5d maximally supersymmetric
Yang-Mills theory (henceforth referred to as $\N=2$ SYM or MSYM) as the worldvolume theory of D4 branes  is of particular interest in the sense that
the latter may be used as a definition of the former
M5 brane theory. The 5d MSYM coupling is related to the circle radius $R$ by
\bea
g_{YM}^2 &=& 4 \pi^2 R
\eea
This 5d MSYM theory involves a solitonic sector of instanton particles corresponding to D0 branes bound to D4 branes, and the required KK spectra along the M-circle (fifth)  direction are correctly reproduced by them \cite{Lambert:2010iw}. Indeed, for the U(1) case, the D4/M5
correspondence has been checked explicitly by direct computation of  partition functions from both sides \cite{Bak:2012ct}.  However the 5d  U(N)
MSYM theory with $N \ge 2$ is certainly perturbatively nonrenormalizable and turns out to involve infinity at six-loop order \cite{Bern:2012di}. Hence
the check of the proposal is not possible in the standard field theoretic framework.

The DLCQ definition of M5 brane theory \cite{Aharony:1997th, Aharony:1997an}
can be of rescue to this situation. (For the effort along the idea of the deconstruction,
see Refs.~\cite{ArkaniHamed:2001ie, Lambert:2012qy,Gustavsson:2012ei}.)
The DLCQ description of $k$ D0-brane sector agrees with
that of the $\N=8$ quantum mechanics (four complex supercharges) on the moduli-space of $k$ instantons based on the ADHM construction of the 5d MSYM theory
\cite{Lambert:2012qy, Lambert:2011gb}.
Our $\N =8$ quantum mechanics below is slightly different in the sense that it involves a potential
arising from turning on the scalar
vacuum expectation values (vev) \cite{Lambert:1999ua}
which introduce another mass scale $\phi_0=\langle \phi^6  \rangle$  to the problem in addition to the
M-circle radius $R$. This $\N =8$ quantum mechanics can be understood from the following DLCQ limit
of the 5d MSYM theory. For the $k$ instanton sector, the corresponding KK momentum is given by
\be
p^5 = \frac{k}{R}
\ee
and the energy by
\be
p^0= E=\sqrt{p_5^2 + {\cal H}_\perp}
\ee
where
\bea
 {\cal H}_\perp &=& {\cal H}_{\rm md} +
O[\, p^2_\perp  (R^2 p^2_\perp)^n(R^2 \phi^2_0)^m ] +O[\, p^2_\perp  (\phi^2_0 p^2_\perp)^n(R^2 \phi^2_0)^m ]\nonumber
\\
 {\cal H}_{\rm md} &=& p^2_\perp +{\cal V}
\eea
with $n, m$ non-negative integers and $n+m \ge 1$.
$p_\perp$ denotes the transverse directional  moduli momentum and ${\cal V}$ is the potential which is of order $\phi^2_0$.
As the $x^5$ direction is circle compactified with the radius $R$, we have the identification
\be
x^5 \ \sim \  x^5 + 2\pi R
\ee
and let us boost the system in the $x^5$ direction with a velocity $u$
\bea
{x'}^{0} =\frac{1}{\sqrt{1-u^2}}(x^0 -u x^5),\ \ {x'}^{5} =\frac{1}{\sqrt{1-u^2}}(x^5 -u x^0), \ \ {x'}^i= x^i
\eea
Let us further introduce $x^\pm =\frac{1}{\sqrt{2}}(x^0 \pm x^5)$, which, under the boost,  transform as
\be
{x'}^{+} =\epsilon x^+ , \ \    \ \           {x'}^{-} =\frac{1}{\epsilon} x^-
\ee
with $\epsilon= \sqrt{(1-u)/(1+u)}$. In the $\epsilon \rightarrow 0$ (infinite momentum) limit, we have the
identification
\be
({x'}^+, \, {x'}^-, \,  {x'}^i) \ \sim \ ({x}^+, \, {x}^- + 2\pi R', \,  {x}^i)
\ee
where $R' = \frac{R}{\sqrt{2} \epsilon}$ and we keep $R'$ finite
by sending also $R$ to zero.
In this DLCQ limit with $k >0$,
\bea
&& {p'}_- = {p'}^+ = \frac{k}{R'} + O(\epsilon^2) \ \rightarrow \  \frac{k}{R'} \\
&&  {p'}_+ = {p'}^- = \frac{R'}{2k} (p^2_\perp +{\cal V}) + O(\epsilon^2) \ \rightarrow \  \frac{R'}{2k} (p^2_\perp +{\cal V})
\eea
while the anti-instanton sector with $k <0$ decouples from the instanton sector  completely because their states become infinitely heavy.
A few comments are in order. The resulting DLCQ Hamiltonian is precisely
 that of the moduli space approximation supplemented by the above mentioned potential term. Its $\N =8 $ supersymmetric
completion is uniquely fixed by the moduli space metric together with the triholomorphic Killing vector $G$
\cite{AlvarezGaume:1983ab,Bak:1999da,Bak:1999vd}, which describes $k$ D0 brane
(Coulomb-branch) dynamics in the presence of $N$ parallel D4 branes.   Due to the potential, we do not have any possible danger
since  the potential is confining asymptotically leading to a finite mass gap in the fluctuation spectra near the instanton
configurations. There are no interactions between different $k$ sectors  as is usual in light-cone frame dynamics. Hence each $k$ sector
of the dynamics can be studied separately. As we shall see below explicitly, this quantum mechanics for a finite $k$ sector
is indeed  well defined and regular.

In this paper, we study the $k=1$,  $\N =8$ quantum mechanics for the gauge group U(N). In order to avoid the singularity
of the modular space geometry, we turn on the spatial (anti-self-dual) noncommutativity.  We adopt then the ADHM construction
\cite{Atiyah:1978ri, Dorey:2002ik}
 of
instanton solutions with general ADHM data. By solving the ADHM constraints explicitly, we shall find the moduli space
for an arbitrary $N$, which corresponds to the Calabi space \cite{Calabi}  times the overall translation $\mathbb{R}^4$. By turning on
the vev of $\phi^6$, the gauge symmetry is broken down to U(1)$^N$. We shall compute the potential explicitly in terms of the
moduli-space coordinates. Thus we claim that the resulting $\N=8$ quantum mechanics describes
the $k=1$ sector of the circle-compactified M5 brane theory in the DLCQ limit.

We test the resulting quantum mechanics by computing the index partition functions for their 1/4-BPS states and
find a perfect agreement with the result in \cite{Kim:2011mv} from the index computation using localization of the 5d MSYM  theory.
These 1/4-BPS states are associated with  the dyonic (electrically-charged) instantons (D0-F1 bound states)
which amount to F-strings stretched between D4 branes in the presence of  instantons \cite{Lambert:1999ua}.
For the minimal $N=2$ case, we are led to the Eguchi-Hanson (EH) space
\cite{Eguchi:1978gw}
times the overall $\mathbb{R}^4$ as the
moduli-space geometry \cite{Lee:2000hp}.
We shall present rather detailed constructions of  related 1/4-BPS states
as well as their supermultiplet structures based on the results in \cite{Bak:1999ip}.

In order to compute the index partition function of the 1/4-BPS states, we use the property of the deformation
invariance of the number of  1/4-BPS states. A detailed account of the relevant index theorem can be found in
\cite{Stern:2000ie}. Here we present a brief account of the index theorem relevant for our discussions. We note that our
$\N=8$ Lagrangian admits a deformation
\be
{\cal L} \ \rightarrow \ \tau^2 {\cal L}\label{resc}
\ee
under which the central charge $Z(\phi_0,Q)$ as a function of only the vev and electric
charges, remains invariant. Furthermore the index
\bea
{\cal I}^+&=& {\mbox{number of selfdual states}} - {\mbox{number of anti-selfdual states}}\label{ind}
\eea
remains invariant. There is no net contribution coming from non-BPS states to the index \cite{Stern:2000ie}. Also it has been made plausible (and was proven for $N=2$ \cite{Stern:2000ie}) that the number of anti-selfdual 1/4-BPS states vanishes, so that the index counts precisely the number of 1/4-BPS states. Now in the limit where $\tau^2$ goes to infinity ($R\rightarrow 0$), the states are localized around the zeroes of
the potential which is non-negative definite. There are $N$ such minima around which the relative moduli-space becomes
$(\mathbb{R}^4)^{N-1}$ which we may interpret as the world volume of $N$ D4 branes where the dyonic instanton Hilbert space at one of these D4 branes has been deleted, henceforth referred to as deleted 
location\footnote{If we compactify one spatial direction of D4 on a circle and T-dualize along that
circle, then what we refer to as deleted location corresponds to the
D3 brane on which a dipole instead of a monopole is located, and whose
dipole charge corresponds to the noncommutativity parameter \cite{Lee:1999xb}.}. In each of the relative $\mathbb{R}^4$, there lives a 4d $\N=8$ superharmonic oscillator, for which we
shall find the 1/4-BPS states explicitly. Due to these $N$ deleted locations, the number of states of one instanton
involving a singly connected F-string scales as $\frac{16}{9}N^3$ for the large $N$, which is unexpected from the
$N^2$ scaling of the MSYM field degrees of freedom.

It should be mentioned that our $\N=8$ quantum mechanics describes not just the 1/4-BPS states but also
generic non-BPS ones of the system. This is contrasted to the index computation in \cite{Kim:2011mv}, where
we do not know the way to deal
with the non-BPS states due to the lack of the formulation.

The paper is organized as follows.
Section 2 presents the 5d U(N) MSYM theory with the noncommutativity turned on. We give the basic properties of the
dyonic instantons which are 1/4-BPS. In Section 3, the $\N=8$ quantum mechanical sigma model \cite{AlvarezGaume:1983ab}
is described. Together with the potential, we set up the BPS equation \cite{Bak:1999da}
whose solutions the 1/4-BPS states
of dyonic instantons. In Section 4, we review the ADHM construction of the instanton moduli space. We obtain the
moduli-space metric explicitly  for the $k=1$ sector leading to the Calabi metric of the relative space. We also compute the
potential as a function of moduli coordinates and identify the triholomorphic Killing vector $G$. In Section 5, we compute
the number of states associated with an instanton involving an F-string singly connected from one D4 to another D4.
This is done adopting the $\tau^2\rightarrow \infty$ deformation of the quantum mechanics leading to $N$ distinct
localization points in the relative space \cite{Lee:1999xb}.
Each of the localized point is characterized by one deleted location of D4 at which no dynamical degrees live. On the other hand,
in each of the remaining D4 branes, the associated part of the moduli-space becomes
$\mathbb{R}^4$, in which lives
one set of  4d $\N=8$ superharmonic oscillator.
Based on this localization, we compute the $k=1$ index partition function of dyonic instantons in Section 6.
Section 7 is devoted for the discussion of the spin and R-symmetries of dyonic instanton states. In Section 8, we take
the EH case of $N=2$ and give the detailed description of states, spin and R-symmetry of the dyonic instanton states based on the results of
\cite{Bak:1999ip}.
We also present the full general treatment of   1/4-BPS states of the 4d $\N=8$ superharmonic oscillator
problem. Based on these results, we compute the more refined version of the index partition function with extra chemical potentials for
the spin ($J^{3+}$) and the R-charges ($R^{3\pm}$) and find a full agreement with the index computation of the 5d MSYM  theory in
\cite{Kim:2011mv}. Finally we show that the counting of states has a smooth commutative limit.
Various technical details as well as some explicit constructions are collected in Appendices.

\section{Five-dimensional $\N=2$ SYM and the dyonic instanton}
We will use 11d notation \cite{Lambert:2010iw} for the 5d $\N=2$ super Yang-Mills (SYM) theory. The classical action is given by
\bea
S &=& \frac{1}{g_{YM}^2} \int dt d^4 x \tr \Bigg( - \frac{1}{4}
 F_{\mu\nu} F^{\mu\nu} - \frac{1}{2} D_{\mu} \phi^{\hat{A}} D^{\mu} \phi^{\hat{A}} + \frac{1}{4} [\phi^{\hat{A}},\, \phi^{\hat{B}}]^2\cr
&&+ \frac{i}{2} \bar{\chi} \Gamma^{\mu} D_{\mu} \chi + \frac{1}{2} \bar{\chi} \Gamma^{\hat{A}}\Gamma_{5} [\phi^{\hat{A}},\chi]\Bigg)
\eea
where
\bea
F_{\mu\nu} &=& \partial_{\mu} A_{\nu} - \partial_{\nu} A_{\mu} - i [A_{\mu},A_{\nu}]
\eea
The spinor $\chi$ is subject to an 11d Majorana and a 6d Weyl condition
\bea
\bar{\chi} &=& \chi^T C\cr
\Gamma_{(6)} \chi &=& - \chi
\eea
where $\Gamma_{(6)} = \Gamma_{012345}$. Our conventions for the gamma matrices are collected in appendix \ref{Spinor}. The on-shell supersymmetry variations read
\bea
\delta \phi^{\hat{A}}&=& i \bar{\omega} \Gamma^{\hat{A}} \chi\cr
\delta A_{\mu} &=& i\bar{\omega} \Gamma_{\mu} \Gamma_5 \chi\cr
\delta \chi &=& \frac{1}{2} \Gamma^{\mu\nu} \Gamma_5 \omega F_{\mu\nu} +
 \Gamma^{\mu} \Gamma_{\hat{A}} \omega D_{\mu} \phi^{\hat{A}} - \frac{i}{2} \Gamma^{{\hat{A}}{\hat{B}}}
\omega [\phi^{\hat{A}},\phi^{\hat{B}}]
\eea
The supersymmetry algebra in a massive dyonic instanton background reads \cite{Lambert:2010iw}
\bea
\{Q,Q^{\dag}\} &=& M - \Gamma^{50} \, \frac{4\pi^2 k}{g_{YM}^2} + \Gamma^{560} Q_E
\eea
Here the central charges are given by
\bea
k &=& \frac{1}{32\pi^2} \int d^4 x \epsilon_{ijkl} \tr \(F_{ij} F_{kl}\)\cr
Q_E &=& \frac{1}{g_{YM}^2} \int_{S^3_{\infty}} d^3\Omega_i \tr \(v^6 F_{0i}\)
\eea
This algebra shows that the dyonic instanton requires a nonvanishing vev $v^6 = \<\phi^6\>$, carries instanton charge $k$, electric charge $Q_E$, preserves 1/4 of SUSY and has the BPS mass $M = |k|/R + |Q_E|$.  There are $12$ broken supercharges out of which $6$ become lowering operators. Acting with these lowering operators on a highest weight state we generate a supermultiplet with $2^6 = 64$ states \cite{Lambert:2010iw}. This analysis does not give us all the 1/4-BPS states though. This is so because we have more fermionic zero modes than broken supercharges, unless the gauge group is $U(1)$ in which case we cannot have any vev and no dyonic instanton. For gauge group $U(2)$ we will obtain $2$ copies of the above $64$-state supermultipliet.

In this paper we will explore the $k=1$ sector of  the dyonic instanton for higher-rank gauge groups U(N).
To find all 1/4-BPS states, we count number of solutions of the $1/4$-BPS equation by transcribing the fermionic zero modes into form-fields on moduli space.

For this analysis we need to regularize the instanton moduli space. We will make a noncommutative deformation
\ben
[x_i,x_j] &=& i\theta_{ij}\label{nc}
\een
where, for the selfdual instanton, we shall assume that $\theta_{ij}$ is antiselfdual. Such a deformation breaks $SO(4) = SU(2)_+ \times SU(2)_b$ rotation symmetry down to $SU(2)_+ \times U(1)_b$. To see this we consider a variation
\bea
\delta^{\pm} x_i &=& \epsilon^{I\pm} \eta_{ij}^{I\pm} x_j
\eea
This gives
\bea
\delta^{\pm} [x_i,x_j] &=& i\epsilon^{I\pm}  [\eta^{I\pm},\theta]_{ij}
\eea
This commutator vanishes for $\delta^{+}$ which means that $SU(2)_+ \subset SO(4)$ is unbroken by this antiselfdual noncommutativity deformation. On the other hand $SU(2)_b$ is broken down to $U(1)_b$.

\section{The $\N=8$ quantum mechanics}
The instanton background preserves $8$ real supercharges. The low-energy dynamics of zero modes of the instanton is therefore described by an $\N=8$ supersymmetric sigma model in one dimension (quantum mechanics) with a potential for the charged or dyonic instanton \cite{Lambert:1999ua}, \cite{AlvarezGaume:1983ab}, \cite{Bak:1999vd}
\bea
S &=& \frac{1}{2}\int dt \Bigg(g_{rs} \(\dot{X}^r \dot{X}^s + i \bar{\psi}^r \gamma^0 D_t \psi^s\) + \frac{1}{6} R_{rstu} \bar{\psi}^r \psi^s \bar{\psi}^t \psi^u\cr
&& - g_{rs} G^{r{\hat{A}}} G^{s{\hat{A}}} - i D_r G^{\hat{A}}_{s} \bar{\psi}^r (\Omega^{\hat{A}} \psi)^s\Bigg)\label{sigmamodel}
\eea
$\N=8$ supersymmetry requires the moduli space metric $g_{rs}$ to be hyper Kahler thus supporting three covariantly constant complex structures $(J^{I-})^r{}_s$. The $G^{r\hat{A}}$ must be triholomorphic and mutually commuting Killing vector fields. The moduli space is on the form
\bea
\M &=& \mb{R}^4 \times \M_{rel}\label{M}
\eea
The three Kahler forms living on this space are on the form
\bea
K^I &=& K^I_{\mb{R}^4} + K^I_{\M_{rel}}
\eea
and the associated complex structures obtained by rising one index by the inverse moduli space metric, are on the form
\bea
J^{I-} &=& \(\begin{array}{cc}
I^I_{\mb{R}^4} & 0\\
0 & I^I_{\M_{rel}}
\end{array}\)
\eea
In later sections when we discuss the relative moduli space we will use the shorter notations $I^{I}$ in place of $I^I_{\M_{rel}}$.  This action describes the dynamics of the moduli parameters $X^r$ and $\psi^r$ (which thus have been given a time dependence) of a dyonic instanton particle. Here $\psi^r$ are two-component Majorana spinors. Despite we have just one time direction here, it is useful to define gamma matrices (associated to an $\N=(4,4)$, $1+1$ dimensional sigma model) as $\gamma^0 = i\sigma^2$, $\gamma^1 = \sigma^1$ and $\gamma = \sigma^3$ where $\sigma^1,\sigma^2,\sigma^3$ denote the $2\times 2$ Pauli sigma matrices. The R symmetry group is $SO(5)$ which rotates $\hat{A}$ as a vector index. If we decompose $\hat{A} = (I,m)$ where $I=1,2,3$ and $m=4,5$, then the $\Omega^{\hat{A}}$ satisfy the half-Clifford algebra of half-gamma matrices
\bea
\{\Omega^I,\Omega^J\} &=& 2\delta^{IJ}\cr
\{\Omega^m,\Omega^n\} &=& -2\delta^{mn}\cr
[\Omega^I,\Omega^m] &=& 0
\eea
where $\Omega^I$ are hermitian and $\Omega^m$ are antihermitian. One gets hermitian generators of $SO(5)$ out of these as follows
\bea
R^{IJ} &=& \frac{i}{4}[\Omega^I,\Omega^J]\cr
R^{mn} &=& -\frac{i}{4}[\Omega^m,\Omega^n]\cr
R^{Im} &=& \frac{i}{2}\Omega^I\Omega^m\label{R}
\eea

One could imagine that we had introduced full hermitian gamma matrices on a doubled space 
\bea
\Gamma^I &=& \Omega^I \otimes \sigma^1\cr
\Gamma^m &=& \Omega^m \otimes (-i\sigma^2)
\eea
Being hermitian, we must take $\Omega^m$ antihermitian. These satisfy the Clifford algebra $\{\Gamma^{\hat{A}},\Gamma^{\hat{B}}\} = 2\delta^{\hat{A}\hat{B}}$. Generators of $SO(5)$ are $K^{\hat{A}\hat{B}} = \frac{i}{4}[\Gamma^{\hat{A}},\Gamma^{\hat{B}}]$ and $K^{IJ} =  R^{IJ} \otimes 1$, $K^{mn} = R^{mn} \otimes 1$ and $K^{Im} = R^{Im} \otimes \sigma^3$.  We then project onto $\sigma^3 = 1$ subspace where we recover the above half-Clifford algebra. An explicit realization is given by
\bea
\Omega^I &=& i(J^{I-})^r{}_s\cr
\Omega^4 &=& i\delta^r_s \gamma^1\cr
\Omega^5 &=& i\delta^r_s \gamma
\eea

The covariant derivative is given by 
\bea
D_t \psi^r &=& \dot{\psi}^r + \Gamma^r_{st} \dot{X}^s \psi^t
\eea
where $\Gamma^r_{st}$ is the Christoffel symbol. Conjugate momenta to $X^r$ are
\bea
p_r &=& g_{rs} \(\dot{X}^s + \frac{i}{2}\Gamma^s_{tu}\bar{\psi}^t \gamma^0 \psi^u\)
\eea
In this paper we will assume that $G^{{\hat{A}}r} = \delta^{{\hat{A}}5} G^r$ which corresponds to one SYM scalar field acquires a vev $\<\phi^6\> = $diag$(v^1,....,v^N)$. In this case the $8$ real supercharges are given by
\bea
Q_{\alpha} &=& \psi^r_{\alpha} p_r + (\gamma^0 \gamma)_{\alpha}{}^{\beta} \psi^r_{\beta} G_r\cr
Q^I_{\alpha} &=& i(J^{I-})^r{}_s \(\psi^s_{\alpha} p_r + (\gamma^0 \gamma)_{\alpha}{}^{\beta} \psi^s_{\beta} G_r\)
\eea
We have the supersymmetry algebra
\bea
\{Q_{\alpha},Q_{\beta}\} &=& 2 \(H \delta_{\alpha\beta} - Z\sigma^1_{\alpha\beta}\)\cr
\{Q^I_{\alpha},Q^J_{\beta}\} &=& 2 \delta^{IJ}  \(H \delta_{\alpha\beta} - Z\sigma^1_{\alpha\beta}\)
\eea
where $H$ is the Hamiltonian and $Z$ is the central charge
\bea
Z &=& G^{r} p_r - \frac{i}{2} D_r G_s \bar{\psi}^r \gamma^0 \psi^s
\eea
We define $4$ complex supercharges 
\bea
\Q &=& \frac{1}{\sqrt{2}} \(Q_1 - i Q_2\)\cr
\Q^I &=& \frac{1}{\sqrt{2}} \(Q^I_1 - i Q^I_2\)
\eea
Also defining $Q^4 = Q$ and letting $i=(I,4)$, the superalgebra generated by them reads
\bea
\{\Q^i,\Q^{j\dag}\} &=& 2\delta^{ij} H\cr
\{\Q^i,\Q^j\} &=& 2i \delta^{ij} Z
\eea
This can be further rewritten as
\bea
\left\{\Q^i \pm i \Q^{i\dag},\(Q^i\pm i \Q^{i\dag}\)^{\dag}\right\} &=& 4\delta^{ij} \(H \pm Z\)\label{susyc}
\eea
Since the left-hand side is non-negative we see that $H \geq |Z|$ where equality holds for BPS saturated states. The condition for a BPS state $\|\Omega\>_+$ which corresponds to the case $Z>0$ reads
\bea
\(\Q^i - i \Q^{i\dag}\) \|\Omega\>_+ &=& 0
\eea
and for an anti-BPS state we have the condition
\bea
\(\Q^i + i \Q^{i\dag}\) \|\Omega\>_- &=& 0
\eea
which corresponds to $Z<0$. If $Z=0$ we require both BPS conditions, which amounts to
\bea
\Q^i \|\Omega\>_0 &=& 0\cr
\Q^{i\dag} \|\Omega\>_0 &=& 0
\eea
and we have no broken supersymmetries in the $\N=8$ sigma model. This case corresponds to a pure instanton.

The fourth supercharge has a particular nice form after transcribing it to form space \cite{Bak:1999da},
\bea
\Q^4 &=& -i \(d - i_G\)
\eea
and the dyonic instanton BPS equation with $Z>0$ becomes
\ben
\Big[(d- i_G)+i (d^\dagger -G)\Big]\Omega=0\label{BPS}
\een
The other BPS equations $\(\Q^I - i \Q^{I\dag}\) \|\Omega\>_+ = 0$ will be automatically satisfied since the mass of the solution saturates the BPS bound. We can then read the equation (\ref{susyc}) backwards. Its left-hand side would have been positive definite had $\(\Q^I - i \Q^{I\dag}\) \|\Omega\>_+$ been non-zero, contradicting the fact that the right-hand side is zero. Therefore solving (\ref{BPS}) will be sufficient.

\section{Brief review of the ADHM construction}
Here we review the ADHM construction of instantons \cite{Atiyah:1978ri, Dorey:2002ik}
which is necessary for our construction of the moduli space metric and the corresponding potential induced by the vev of the scalar field.

The basic object for the ADHM constraint is the $(N+ 2k)\times 2k$ complex-valued matrix $\Delta_{\lambda m \dot{\alpha}}$, which is assumed to be
linear in the 4d spatial coordinates $x_i$ ($i,j,\cdots =1,2,3,4$). Only in this section we will assume a generic instanton number $k$ and let the instanton indices $m, n,\cdots$ run over $1,2, \cdots k$. Later on we will fix $k=1$. The indices $\lambda, \mu, \cdots=1,2, \cdots, N+2k$ are decomposed as $u\oplus m{\alpha}$, $v\oplus n{\beta}$, $\cdots$ with
$u,v,\cdots =1,2,\cdots, N$. We use the notation $\bar{\Delta}^{m \dot{\alpha},\lambda} =(\Delta_{\lambda, m \dot{\alpha}})^*$. Then $\Delta$ can be parametrized as
\be
\Delta_{\lambda, n\dot{\alpha}}= \Delta_{u\oplus m{\alpha},n\dot{\alpha}}
=\left(\begin{array}{c}
w_{u n \dot{\alpha}}\\  (X_{imn}+{x}_i \delta_{mn}) \bar{q}_{i\alpha\dot\alpha}
\end{array}\right)
\ee
where $\bar{q}_{i\alpha\dot\alpha}$ are as specified in Eq (\ref{quaternions}). We shall require the ADHM constraint
\be
\bar{\Delta}^{m\dot{\alpha}, \lambda}\Delta_{\lambda, n\dot{\beta}} = \delta^{\dot{\alpha}}_{\dot{\beta}}(f^{-1})^{m}{}_{n}
\ee
where $f$ is an  $x$-dependent $k\times k$ Hermitian matrix. To get the instanton solution,
one introduces an $(N+2k)\times N$ matrix $U_{\lambda u}$ satisfying
\be
\bar\Delta^{m\dot{\alpha},\lambda} U_{\lambda u} =0\,, \ \ \ \bar{U}^{u\lambda} U_{\lambda v}= \delta^u_v
\ee
Then the gauge field is given by
\be
(A_i)^u{}_v =  i \bar{U}^{u\lambda} \partial_i U_{\lambda v}
\ee
whose field strength is self-dual ($F= *_4 F$).
We
choose noncommutativity parameter defined by (\ref{nc}) as $\theta_{ij}= \zeta \eta^{3-}_{ij}$ as suitable for selfdual instantons. Then the ADHM constraint becomes
\ben
0&=&\bar{w}^{m\dot{\alpha},u}{w}_{u,n\dot{\beta}} (\sigma^I)^{\dot\beta} \,_{\dot\alpha}+i
 [X_i+{x}_i, X_j+ {x}_j]^m{}_n \eta^{I-}_{ij}\nonumber \\
&=&  \bar{w}^{m\dot{\alpha},u}{w}_{u,n\dot{\beta}} (\sigma^I)^{\dot\beta} \,_{\dot\alpha}+i
 [X_i, X_j]^m{}_n \eta^{I-}_{ij}-4 \zeta \delta^m_n \delta^{I3}
\label{ADHMconstraints}
\een
together with $X^\dagger_i= X_i$.

\subsection{Moduli space metric}
The moduli space metric can now be computed starting from the flat metric
\ben
ds^2 &=& \tr_k \(d\bar{w}^{\dot\alpha} dw_{\dot\alpha} + dX^i dX^i\)\label{flat1}
\een
by imposing the ADHM constraint and an appropriate $U(k)$ gauge fixing condition. We will clarify this construction in section \ref{calabi} where we obtain the moduli space metric for $k=1$.

\subsection{Potential}
The scalar field equation in the instanton background
\be
D_i D_i \phi=0
\ee
can be solved for any given ADHM data. The solution is given by (see for instance the appendix in Ref \cite{Dorey:2002ik})
\be
\phi= \bar{U}{\cal J}U= \bar{U}\left(
\begin{array}{cc}
\phi_0 & 0 \\
0 & \varphi\, I_{2\times 2}
\end{array}
\right)   U
\ee
where $\phi_0$ is the vev of the scalar field (which we will choose as an $N\times N$ diagonal matrix) and $\varphi$ is
the $k\times k$ $x$-independent Hermitian matrix satisfying
\be
[X_i,[X_i,\varphi]]+\frac{1}{2}\Big( \bar{w}^{\dot{\alpha}} {w}_{\dot{\alpha}} \varphi +\varphi \,
\bar{w}^{\dot{\alpha}} {w}_{\dot{\alpha}}  \Big )=\bar{w}^{\dot{\alpha}} \phi_0 {w}_{\dot{\alpha}}
\ee
The potential of the ${\cal N}=8$ supersymmetric quantum mechanics can then be obtained by evaluating
\be
V= \frac{1}{2g^2} \int d^4 x {\rm tr} D_i \phi D_i \phi = \frac{1}{2g^2} \int_{S^3_\infty} 
d^3\Omega_i \, 
{\rm tr}\,\phi D_i \phi
\ee
With a short computation, one has
\be
D_i \phi = -\bar{U}(\partial_i \Delta) f \bar{\Delta} U  -\bar{U} \Delta f (\partial_i \bar{\Delta}) U
\ee
and
\be
{x^i} D_i \phi \rightarrow \frac{1}{ x^2} \Big(
\phi_0  w_{\dot{\alpha}}\bar{w}^{\dot{\alpha}}+ w_{\dot{\alpha}}\bar{w}^{\dot{\alpha}}\phi_0  -2
w_{\dot{\alpha}}\varphi \bar{w}^{\dot{\alpha}}
\Big)
\ee
leading to the potential
\be
V= \frac{2\pi^2}{g^2} {\rm tr} \Big( \bar{w}^{\dot{\alpha}}
\phi^2_0  w_{\dot{\alpha}}-\bar{w}^{\dot{\alpha}}
\phi_0  w_{\dot{\alpha}}\varphi
\Big)
\ee

\section{Calabi metric from ADHM constraints}\label{calabi}
Let us now consider $k=1$ of U(N) noncommutative instanton problem. Since the center-of-mass part of the metric decouples,
we shall set $X_i=0$ and consider only the relative part. Our starting point is the flat metric on $\mb{H}^{N} = \mb{R}^{4N}$. We map $4N$ real coordinates $y^i_{u}$ ($u=1,\cdots,N$) into $N$ quaternionic coordinates
\bea
y_u &=& y^i_{u} q_i
\eea
using the quaternion basis (\ref{quaternions}). The flat metric on $\mb{H}^{N}$ can be expressed as
\be
ds^2= \sum^N_{u=1}dy_u d\bar{y}_u
\ee
Here we suppress the overall coefficient of this metric, which is given by $\frac{k}{R}= \frac{4\pi^2}{g^2}$ with $k=1$.\footnote{Our convention for the Lagrangian for the SUSY quantum mechanics is given in (\ref{sigmamodel}) and this fixes the normalization of the moduli space metric $g_{rs}$.} By introducing the Hopf map $\mb{H}^N \rightarrow \mb{R}^{3N}$,
\bea
y_u q_3 \bar{y}_u &=& 4 x^I_{u} q_I
\eea
the above flat metric takes the form \cite{Gibbons:1996nt}
\be
\sum^N_{u=1}\Big(
C_u d\vec{x}^2_u +C_u^{-1} \sigma^2_{\psi_u}
\Big)
\label{metric1}
\ee
Here $\vec{x}_u$ refers to $x^I_u$ and
\bea
C_u &=& \frac{1}{x_u}
\eea
with $x_u = \sqrt{\vec{x}_u^2}$. Associated with the circle-fiber over $\mb{R}^{3N}$ we define
\bea
\sigma_{\psi_u} &=& d\psi_u + A_u\label{flat2}
\eea
The $4\pi$-periodic angles $\psi_u$ are defined from the quaternions by
\bea
y_u &=& a_u \exp \(q_3 \frac{\psi_u}{2}\)
\eea
with $\bar{a}_u = -a_u$ purely imaginary. The vector potentials $A_u$ are related to the functions $C_u$ as
\bea
*dA_u &=& dC_u
\eea
If we parametrize
\bea
\x_u &=& x_u(\sin \theta_u \cos \phi_u,\sin \theta_u \sin \phi_u,\cos \theta_u)
\eea
where the coordinates $(\theta_u,\phi_u)$ are the usual polar coordinates on $S^2$ base manifold, then we have
\bea
A_u &=& (1+\cos \theta_u)d\phi_u
\eea
In the original Cartesian coordinates we find that
\bea
A_u &=& \frac{1}{x_u \(x_u - x^3_{u}\)}\(x^1_u dx^2_u - x^2_u dx^1_u\)
\eea
We present a derivation of this form of the flat metric in the appendix \ref{Flat}.

The ADHM constraints (\ref{ADHMconstraints}) are expressed in terms of $w_{u\dot{\alpha}} \in \mb{C}^{2N}$. Therefore we wish to have a map $\mb{H}^N \rightarrow \mb{C}^{2N}$. In the $2\times 2$ realization of quaternions we find that
\bea
\bar{y}_u &=& y^{\dag}_u
\eea
and
\bea
y_u &=& \(\begin{array}{cc}
w_{u\1} & -\bar{w}^{\2 u}\\
w_{u\2} & \bar{w}^{\1 u}
\end{array}\)
\eea
where we define
\bea
w_{u\1} &=& y_{u4} - i y_{u3}\cr
w_{u\2} &=& y_{u2} - i y_{u1}\cr
\bar{w}^{\1 u} &=& y_{u4} + i y_{u3}\cr
\bar{w}^{\2 u} &=& y_{u2} + i y_{u1}
\eea
This now defines a map $\mb{H}^N \rightarrow \mb{C}^{2N}$.

The ADHM constraints
\bea
\sum_u w_{u\dot\alpha} (\sigma_I)^{\dot\alpha}{}_{\dot\beta} \bar{w}^{\dot\beta u} &=& 4 \zeta_I
\eea
can now be written as
\bea
\sum_u y_u q_3 \bar{y}_u &=& 4 \zeta_I q_{I}
\eea
and can be solved as
\be
\vec{x}_N= -\sum^{N-1}_{A=1} \vec{x}_A +  \zeta \hat{e}_3\label{ADHMC}
\ee
Our indices range as $u,v,\cdots=1,\cdots,N$ and $A,B,\cdots=1,\cdots,N-1$ respectively, and $\hat{e}_3 = (0,0,1)$. We insert this into (\ref{metric1}) to eliminate $\vec{x}_N$.  Furthermore we have the U(1) symmetry
\be
y_u \rightarrow y_u \exp\(q_3 t\)
\ee
which acts as a translation of the angles
\be
\psi_u \rightarrow \psi_u +2t
\ee
Since we shall mod out this $U(1)$ symmetry, we introduce $U(1)$ invariant angles
\be
\varphi_A= \psi_A -\psi_N
\ee
and define corresponding one-forms
\be
\sigma_{\varphi_A}=d\varphi_A +A_A-A_N\label{varphi}
\ee
As an intermediate step in obtaining the metric, we compute
\bea
\sum_u \frac{1}{x_u}d\x_u^2 &=& \sum_A \frac{1}{x_A} d\x_A^2 + \sum_{A,B} \frac{1}{x_N} d\x_A \cdot d\x_B\cr
\sum_u x_u \sigma_{\psi u}^2 &=& \sum_A x_A \(\sigma_{\varphi_A}^2 - \frac{1}{L} \(x_A \sigma_{\varphi_A}\)^2\) \cr
&&+ L \(\sigma_{\psi_N} + \frac{1}{L} x_A \sigma_{\varphi_A}\)^2
\eea
where we define
\ben
L &=& \sum_u x_u\label{L}
\een
We now mod out by the U(1) gauge symmetry by putting the momentum conjugate to $\psi_N$ to zero. This kills the last term. The resulting metric is the Calabi space metric \cite{Calabi,Hatzinikitas:1998dc,Kim:2001kp,Lee:1999xb}
\bea
ds^2 &=& C_{AB} d\x_A \cdot d\x_B + C^{-1}_{AB} \sigma_{A} \sigma_{B}
\eea
where
\ben
C_{AB} &=& \frac{\delta_{AB}}{x_A} + \frac{1}{x_N}\cr
C^{-1}_{AB} &=& x_A \delta_{AB} - \frac{1}{L} x_A x_B\label{C}
\een

\section{The potential for the Calabi space}
We take the scalar vev
\be
\phi_0= 
{\rm diag}[v_1,v_2, \cdots, v_N]
\ee
For the $k=1$ case, the scalar data $\varphi$ is given by
\be
\varphi=  \frac{\bar{w}^{\dot{\alpha}} \phi_0 {w}_{\dot{\alpha}}}{\bar{w}^{\dot{\alpha}} {w}_{\dot{\alpha}}}
\ee
Then the  potential becomes
\be
V=\frac{2\pi^2}{g^2} \Big( \bar{w}^{\dot{\alpha}}
\phi^2_0  w_{\dot{\alpha}}-\frac{(\bar{w}^{\dot{\alpha}}
\phi_0  w_{\dot{\alpha}})^2}{\bar{w}^{\dot{\alpha}} {w}_{\dot{\alpha}}}
\Big)
\ee
Noting
\be
{\bar{w}^{\dot{\alpha}} {w}_{\dot{\alpha}}} = 4L
\ee
which follows from (\ref{L}) together with the usual relation between the radii of spheres,
$4 x_u = w_{u\dot{\alpha}} \bar{w}^{\dot{\alpha}u}$ in the Hopf map $S^3 \mapsto S^2$, this is evaluated as
\be
V=\frac{2\pi^2}{g^2}\,4 
\Big(
\sum_{u=1} x_u v_u^2 - \frac{1}{L}\big(\sum_{u=1} x_u v_u \big)^2
\Big)
\ee
One may rewrite this as
\be
V=\frac{2\pi^2}{g^2} \frac{4}{L} \sum_{u < v} x_u x_v (v_u  -v_v)^2 = \frac{2\pi^2}{g^2}
C^{AB}2(v_A-v_N)2(v_B-v_N)
\ee
Hence one can see that each Killing direction $\varphi_A$ is weighted by $2(v_A-v_N)$, which corresponds to an F-string (W-boson) connecting
D4$_N$ to D4$_A$.
When  $v_u -v_v\neq 0$ for any $u\neq v$, the U(N) gauge symmetry is maximally broken down to U(1)$^N$. For this case,
one finds that the potential  is non degenerate near any zeroes of the potential  and  receives the nontrivial quadratic contributions.
Finally the corresponding Killing vector  $G$ can be identified as
\be
G= \sum^{N-1}_{A} 2(v_A-v_N) \frac{\partial}{\partial \varphi_A}\label{G}
\ee
The electric charge $Q_A$ is defined by
\be
Q_A = -2 i {\cal L}_{\partial_{\varphi_A} }\ \in \ {\mathbb{Z}}
\ee
while $Q_N = -\sum_A Q_A$  due to the overall U(1) invariance of the Calabi metric.

\section{Localization to $\mathbb{R}^{4(N-1)}$ and counting of 1/4-BPS states}
We now come to the central part of this paper. We will count the number of 1/4-BPS states for U(N) gauge group in the sector with instanton number $k=1$. In subsequent sections we will also classify these states according to their representations of the unbroken global symmetry group ${\cal G}=SU(2) \times SO(4)$ (this symmetry group will be explained in great detail in
subsequent sections. Let us for now only mention that ${\cal G}$
corresponds to unbroken Lorentz times R-symmetries) of the underlying 5d SYM theory. We have not
succeeded to find exact solutions to the relevant 1/4-BPS equation (\ref{BPS}), not even for the
simplest case when $N=2$. (A vev when $N=1$ has no significance
so there would be no 1/4-BPS states in that case.) Instead we will make use of the index (\ref{ind}) to count the number of 1/4-BPS states. A detailed account
of this index can be found in Ref.~\cite{Stern:2000ie}. The index is invariant under the rescaling (\ref{resc}), which
allows us to localize to points of minima of the potential
where the potential is that of an $\N=8$ supersymmetric harmonic oscillator, and the moduli space
metric is locally flat and on the form $\mb{R}^{4(N-1)}$. Furthermore, it will be sufficient to solve this
BPS equation in $\mb{R}^4$ (corresponding to taking $N=2$) due to a factorization property of the harmonic oscillator.
This BPS equation and its solutions have been obtained in \cite{Stern:2000ie} by viewing $\mb{R}^4 = \mb{C}^2$.
However for our purpose of classifying these BPS states according to their representations of ${\cal G}$ we find it more
convenient to obtain these solutions in a vielbein basis which is constructed out of the
 Maurer-Cartan forms on $S^3 = SU(2)$ and hence our view is that $\mb{R}^4 = \mb{R}_+ \times S^3$.
We present this BPS equation along with detailed steps on how to obtain its solutions in appendix \ref{Instanton}.

Let us now describe how the Calabi metric near any of the minima of the potential becomes flat $\mb{R}^{4(N-1)}$. The Calabi metric can be expressed as
\bea
ds^2 &=& \sum_{u=1}^{N} \frac{1}{x_u} d\x_u^2 + \frac{1}{x_1+\cdots +x_N} \sum_{u>v=1}^N x_u x_v (\sigma_u-\sigma_v)^2
\eea
where we define
\bea
\sigma_N &\equiv & 0
\eea
and we assume that
\bea
\x_1 + \cdots + \x_N &=& \vec{\zeta}
\eea
The central charge is given by
\bea
G &=& \sum_{u=1}^{N} v^{uN} Q_u
\eea
Minima of the potential are uniquely characterized by specifying a D4 brane $u_0$ that we refer to as deleted location (see also \cite{Lee:1999xb}). We thus specify $u_0 = 1,\cdots,N$ and
take $x_{u_0} = \zeta$ while all other $x_u = 0$ ($u\neq u_0$). The metric near the minimum with a deleted location at $u_0$ is given by
\bea
ds^2 &=& \sum_{u\neq u_0} \frac{1}{x_u} d\x_u^2 + 
\sum_{u\neq u_0} x_u (\sigma_u - \sigma_{u_0})^2
\eea
This metric is flat and describes the space $\mb{R}^{4(N-1)}$. We identify this part of moduli space
with space of $N-1$ out of $N$ D4 branes as illustrated in Figure 1.

\begin{figure}\label{fig1}
\begin{center}
\includegraphics[scale=0.9]{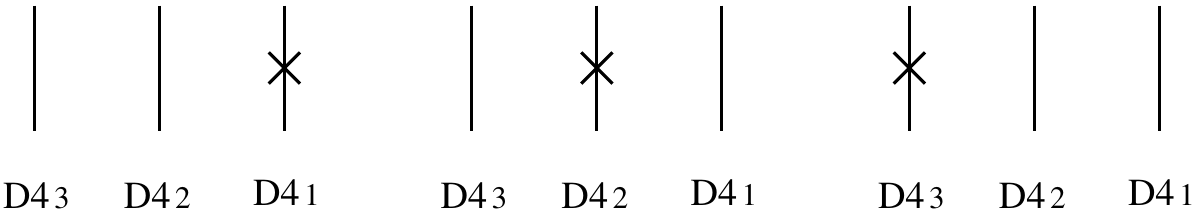}
\end{center}
\caption{\small We illustrate D4 brane configurations with various deleted locations for $N=3$.}
\end{figure}

The U(1)$^{N-1}$ angles $\varphi_u$, where we define $\varphi_N = 0$, sit in the metric as
\bea
\sum_{u=1}^N (d\varphi_u - d\varphi_{u_0})^2
\eea
which motivates us to define local $\t U(1)^{N-1}$ angles
\bea
\t \varphi_u &=& \varphi_u - \varphi_{u_0}
\eea
where apparently $\t \varphi_{u_0} = 0$. The associated charges are related as
\bea
Q_u &=& \t Q_u\,, \quad u \neq u_0\cr
Q_{u_0} &=& - \sum_{u\neq u_0}^N \t Q_u
\eea
The relation (viewed as a map from $N-1$ charges into $N-1$ charges) can be inverted as
\bea
\t Q_A &=& Q_A\,, \quad A \neq u_0\cr
\t Q_N &=& -\sum_{A=1}^{N-1} Q_A
\eea
for $u_0 =1,\cdots,N-1$. If $u_0 = N$,
we have
\bea
\t Q_A &=& Q_A
\eea

The central charge can be expressed as
\bea
G &=& \sum_{u=1}^N (v^u - v^{u_0}) \t Q_u
\eea
in terms of local charges.

In Ref.~\cite{Stern:2000ie} was obtained the number of $1/4$-BPS states.
We present another derivation of this result as well as further details on representations of these states in section \ref{internal}. These studies show that as factorized 4d superharmonic oscillators, labeled by $u\neq u_0$, one has the following number of BPS states at each such $u$:
\ben
n_u &=& \left\{\begin{array}{cl}
4|Q_u| & {\rm if}\  \(v^u - v^{u_0}\) Q_u \ >\ 0 \nonumber \\
1    &   {\rm if}\ v^u - v^{u_0} \neq 0 \ {\rm and } \ Q_u=0 \nonumber \\
 0  & {\rm otherwise}
\end{array}\right.
\een
and the total number of 1/4-BPS states is given by the product
\bea
n &=& \prod_{u\neq u_0} n_u
\eea
We are interested in the case of one connected F-string stretching from
D4$_v$ to D4$_u$  (which we denote by F$_{uv}$) where $1\leq u<v \leq N$ and we may order the
branes such that \footnote{This ordering of vev is always possible by utilizing a group element of the U(N) gauge symmetry which is a permutation.}
\bea
v_1 > v_2 > \cdots > v_N
\eea
Such a string is associated with charges
\bea
q_a &=& 1 \quad (a=u,\cdots,v-1)\cr
q_a &=& 0 \quad {\mbox{otherwise}}
\eea
The charge assignment of $Q_u$ can be understood from the caloron picture. An alternative derivation of the
Calabi metric starting from the caloron dynamics is presented in Appendix C and,  there, the relation between $Q_u$
and the string charges $q_a$ is explained in detail. We present here simply the result: the charges are related by
\bea
Q_1 &=& q_1 - q_N\cr
Q_2 &=& q_2 - q_1\cr
\vdots \cr
Q_{N} &=& q_{N} - q_{N-1}
\eea
where only $N-1$ of these charges are independent due to the constraint $\sum_u Q_u = 0$. The $q_A$ counts
the number of F1$_{A+1,A}$-strings. $q_N$ vanishes in the decompactification limit of the caloron configurations.

In our case of
an F$_{uv}$-string stretched between D4$_v$ and D4$_u$, 
we now find that
\bea
Q_{u} &=& 1\cr
Q_{v} &=& -1
\eea
and $Q_w = 0$ for $w \neq u,v$. These charge assignments are illustrated in Figure 2.
\begin{figure}\label{fig2}
\begin{center}
\includegraphics[scale=0.9]{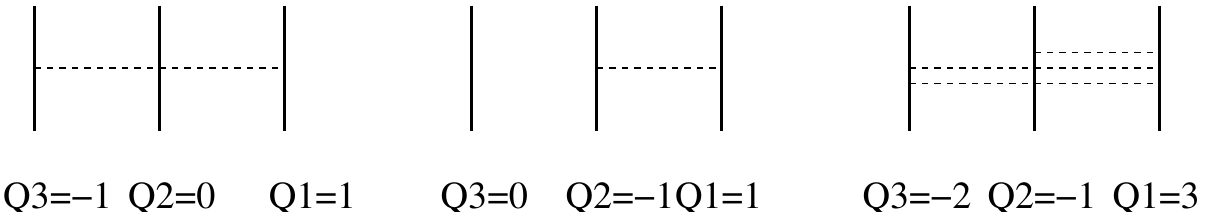}
\end{center}
\caption{\small Various configurations of F-strings
stretched between D4 branes and corresponding charges are illustrated for $N=3$.}
\end{figure}

We now map this to charges as seen by the local flat metric near the minima. Let us first assume that $u=1$ and $v=N$. Then we have $Q_1 = 1$, $Q_N = -1$ and all the other charges are zero. Then
\bea
\t Q_1 &=& 1\cr
\t Q_N &=& -1
\eea
if $u_0 \neq \{1,N\}$. If $u_0 = 1$ then $\t Q_1$ is not defined and the only nonvanishing charge is $\t Q_N = Q_N = -1$. If $u_0 = N$ then $\t Q_N$ drops out and the only nonvanishing charge is $\t Q_1 = 1$. If $u_0 = 1$ then the central charge is
\bea
G &=& \(v^N - v^1\) \t Q_N
\eea
which is positive so this yields $4$ BPS states. If $u_0 = N$ then the central charge is
\bea
G &=& \(v^1 - v^N\) \t Q_1
\eea
which is positive, so this yields $4$ BPS states. Let us now take $u_0 = 2,\cdots,N-1$. Then the central charge is
\bea
G &=& \(v^1 - v^{u_0}\) \t Q_1 + \(v^N - v^{u_0}\) \t Q_N
\eea
Both terms are positive so we find $4\times 4 = 16$ BPS states.

Let us now assume that $1<u<v< N$. Then $q_u = \cdots  = q_{v-1} = 0$ and the rest is vanishing. Then
\bea
Q_u &=& 1\cr
Q_v &=& -1
\eea
For $u_0\neq u,v$ and we get
\bea
\t Q_u &=& 1\cr
\t Q_v &=& -1
\eea
and all other charges are zero, including $\t Q_N$. If $u_0$ is not at the boundary of the F-string, then the central charge is
\bea
G &=& (v^{u} - v^{u_0}) \t Q_{u} + (v^{v} - v^{u_0}) \t Q_{v}
\eea
This is positive if $u < u_0 <v$. Otherwise the two terms have opposite sign and we get no BPS states, unless $u_0$ is at the one boundary, say $u_0 = u$ and then $\t Q_u$ gets absent while we get $\t Q_N = 0$ and so the only non-vanishing charge is $\t Q_v = -1$. The central charge is
\bea
G &=& (v^v - v^u)\t Q_v
\eea
and this is positive and so we find $4$ BPS states. For the other boundary, $u_0 = v$ we find $\t Q_u = 1$ as the only non-vanishing charge. The central charge is
\bea
G &=& (v^u - v^v)\t Q_{u}
\eea
which is again positive and so we find $4$ BPS states.

Let us summarize our findings. If an F$_{1N}$-string is connected from D4$_N$ to D4$_1$, the number of BPS states becomes
\be
4 \times 2 + 16 \times (N-2)
\ee
where $8$ comes from the two boundary deleted locations whereas the $16(N-2)$ comes from the contributions of the internal deleted locations.
If an F$_{uv}$-string is connecting D4$_u$ to D4$_v$ with $u < v$, the number of BPS states is
\be
4 \times 2 +16 \times (v-u -1)
\ee
where $4\times 2$ comes from the deleted locations at the boundaries D4$_u$ or D4$_v$ and
$16\times (v-u-1)$ comes from the internal deleted locations.
If the deleted locations are located at D4 branes outside F$_{uv}$-string, one does not have any BPS states.

We have dyonic instantons which correspond to F$_{uv}$-strings with deleted locations at either one of the two boundary D4 branes of the string. There are
\bea
2 \times 4\,\,\frac{N(N-1)}{2}
\eea
such dyonic instanton states. We also have dyonic instantons which correspond to F$_{uv}$-strings with their deleted location at an internal D4 brane.
The number of such states is
\bea
4 \times 4\,\, \frac{N(N-1)(N-2)}{6}
\eea

\section{The one-instanton partition function}
In the brane picture we have $N$ separated D4 branes with separations
\bea
v_{uv} &=& v_u - v_v
\eea
where again  $v_1>v_2>\cdots >v_N$. We select an index $u_0=1,\cdots, N$ and a corresponding brane D4$_{u_0}$. This
 brane is distinguished by that no $1/4$-BPS states can be located at this brane irrespectively how the F-strings are being stretched.
The instanton partition function is given by the sum over deleted locations
\bea
Z_N &=& \sum_{u_0=1}^N Z_{u_0,N}
\eea
Assume that we have an F$_{uv}$-string that stretches between D4$_u$ and D4$_v$. A deleted location $u_0$ can be classified into three types: internal if $u<u_0<v$, boundary if $w=u$ or $w=v$, and exterior otherwise. In Figure 3 we illustrate two boundary deleted locations and one internal deleted location for
an F$_{13}$-string connecting D4$_1$ to D4$_3$.

\begin{figure}\label{fig3}
\begin{center}
\includegraphics[scale=0.9]{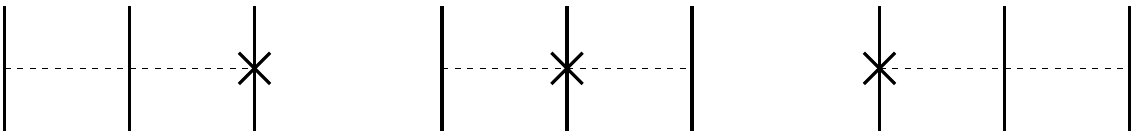}
\end{center}
\caption{\small An F$_{13}$-string connecting D4$_1$ to D4$_3$ with three deleted locations is illustrated.}
\end{figure}

We will proceed by induction. When $N=2$ we have two boundary deleted locations only. From any one of these boundary
deleted locations we have the contribution
\bea
Z_2 = 1 + \sum_{n=1}^{\infty} 4n e^{-\beta nv_{12}} = \coth^2 \frac{\beta v_{12}}{2}
\eea
Here $n$ counts the number of $F$-strings stretching between D3$_1$ and D3$_2$. In the exponent we have the central charge or the BPS energy times a parameter $\beta$. These strings are BPS and the energies add up so that $n$ $F$-strings have the energy $nv_{12}$. The degeneracy of a state of energy $nv_{12}$ is $4Q_1$ if the charge given by $Q_1 = q_1 - q_2 = n - 0$ is positive. If the charge vanishes, $Q_1 = 0$, we have instead degeneracy $1$ and we have $n=0$ and energy $E_{n=0} = 0$. This state gives the contribution $1$ to the instanton partition function $Z_2$.

For general $N$,
the central charge is
\bea
G = \sum_{A=1}^{N-1} v_{AN} Q_A = \sum_{u} v_{uu_0} \t Q_u
\eea
The state is BPS only if
\bea
v_{uu_0} \t Q_u \geq 0
\eea
for each $u$. Since $v_{uu_0} > 0$ for $u=1,\cdots,u_0-1$ and $v_{uu_0} < 0$ for $u=u_0+1,\cdots,N$, this condition is equivalent with
\bea
0 \leq q_1 \leq q_2 \leq \cdots \leq q_{u_0-1}\cr
q_{u_0} \geq q_{u_0+1} \geq \cdots \geq q_N = 0
\eea
We illustrate this BPS condition in Figure 4.
\begin{figure}\label{fig4}
\begin{center}
\includegraphics[scale=1.0]{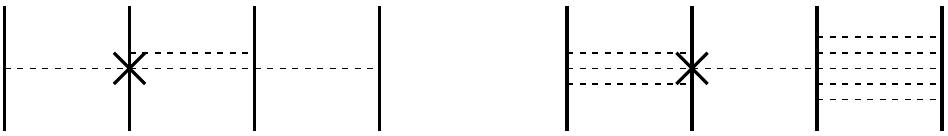}
\end{center}
\caption{\small Left picture is a BPS configuration in which the chages $q_u$ do not increase in the both directions away from  the deleted location.
Right picture is a non-BPS configuration since the chages $q_u$  increase at least once away from  the deleted location. .}
\end{figure}

Let us assume that $N=3$. For the various possible deleted locations at $u_0=1,2,3$ respectively, we find the potential is given by
\bea
G(u_0=1) &=& v_{21} \t Q_2 + v_{31} \t Q_3\cr
G(u_0=2) &=& v_{12} \t Q_1 + v_{32} \t Q_3\cr
G(u_0=3) &=& v_{13} \t Q_1 + v_{23} \t Q_2
\eea
We are only interested in BPS states. For $u_0 =1$ this means
\bea
\t Q_2 &=& -m\cr
\t Q_3 &=& -n
\eea
For $u_0 =2$ this means
\bea
\t Q_1 &=& m\cr
\t Q_3 &=& -n
\eea
and for $u_0=3$ this means
\bea
\t Q_1 &=& m\cr
\t Q_2 &=& n
\eea
Here $m,n=0,1,2,\cdots$. The partition function is
\bea
\coth^2 \frac{\beta v_{12}}{2} \coth^2 \frac{\beta v_{13}}{2} + \coth^2 \frac{\beta v_{12}}{2} \coth^2 \frac{\beta v_{23}}{2} + \coth^2 \frac{\beta v_{13}}{2} \coth^2 \frac{\beta v_{23}}{2}
\eea
where each term corresponds to $u_0 = 1,2,3$ respectively. To see this, we use that for any given deleted location, if $m=n=0$ we have one state. For $m=0$ and $n>0$ we have $1 \times 4n$ states. For $m>0$ and $n=0$ we have $4m \times 1$ states. For $m>0$ and $n>0$ we have $4m \times 4n$ states.

To obtain the partition function of higher $N$, let us first assume the deleted location is on the first brane $u_0 = 1$ and let us denote the partition function at $N$ by $Z_N^1$. Then add an $(N+1)$-th brane at $v_{N+1}$. The corresponding partition function will then become
\bea
Z^1_{N+1} &=& Z^1_N \coth^2 \frac{\beta v_{1,N+1}}{2}
\eea
We also know that for $N=2$ we have
\bea
Z^1_2 &=& \coth^2 \frac{\beta v_{12}}{2}
\eea
The recursion relation can now be solved with this boundary condition as
\bea
Z^1_N &=& \prod_{u=2}^N \coth^2 \frac{\beta v_{1,u}}{2}
\eea
By reflection symmetry we also deduce that if the delocation point is $u_0 = N$, then
\bea
Z^N_N &=& \prod_{u=1}^{N-1} \coth^2 \frac{\beta v_{u,N}}{2}
\eea
We proceed by induction to compute $Z_{N+1}$. Let us assume the partition function for $N$ is known and given by
\bea
Z_N &=& \sum_{u_0=1}^N Z^{u_0}_N
\eea
By adding an $(N+1)$-th brane at $v_{N+1}$, we find that
\bea
Z^{u_0}_{N+1} &=& Z^{u_0}_N \coth^2 \frac{\beta v_{u_0,N+1}}{2}
\eea
for $u_0 = 1,\cdots,N$. If the deleted location is placed on the brane $u_0 = N+1$ we get the contribution
\bea
\prod_{v=1}^N \coth^2 \frac{\beta u_{v,N+1}}{2}
\eea
In summary we find
\bea
Z_{N+1} &=& \sum_{u_0 =1}^N Z^{u_0}_N \coth^2 \frac{\beta v_{u_0,N+1}}{2} + \prod_{v=1}^N \coth^2 \frac{\beta u_{v,N+1}}{2}
\eea
This recursion relation with given boundary condition is uniquely solved by
\bea
Z_N &=& \sum_{u=1}^N \prod_{v\neq u} \coth^2 \frac{\beta v_{uv}}{2}
\eea
Let us now compare this to the result that was obtained in \cite{Kim:2011mv}. In this reference a generalized Witten
index was computed. By specializing this to the one-instanton $k=1$ sector and by choosing
parameters appropriately\footnote{As explained in \cite{Kim:2011mv} this amounts in the notations
of this reference to taking $\gamma_2 = \pi$ in order to cancel the $(-1)^F$ factor in their index. Also we
shall take $\gamma_R = 0$.}, we can descend to the quantity
\bea
\tr_{k=1,{\mbox{1/4-BPS states}}}\exp \(-\beta (H - v_u Q_u) - \mu_u Q_u\)
\eea
We can furthermore bring this into a partiton function over 1/4-BPS states, $\tr_{k=1, 1/4 \, {\rm BPS}}\, \exp \(- \beta H\)$, by taking
\bea
\mu_u &=& \beta v_u
\eea
which leads to a perfect match with our partition function $Z_N$.

Let us comment that the two methods to compute this
partition function is very different. While we rely on moduli space of dyonic instantons, Ref.~\cite{Kim:2011mv} makes no use of this moduli space.

We can consider a refined index where the spin content of the 1/4-BPS states is taken into account by including chemical potentials. We compute the refined index in section \ref{refined} but before that we need to understand the spin content of our 1/4-BPS states.

\section{Spin and R-symmetry representations}
We will now obtain the $SU(2)_+ \times SO(4)$ representations of the $1/4$-BPS states we have constructed as $p$-forms on moduli space. Since the moduli space is on the form of (\ref{M}) we can study this problem for each factor of moduli space separately.

\subsection{Center of mass $\mb{R}^4$ part of moduli space}
We let $\delta_r A_i$ denote bosonic zero modes where $r = 1,\cdots,$ dim$\M_N$ is a curved index on the one-instanton moduli space $\M_N$ and $i=1,2,3,4$ is a spatial index of 5d SYM. We can then express the fermionic zero modes $\chi$ as
\bea
\chi &=& \delta_r A_i \Gamma_i \E_+ \psi^r
\eea
where $\psi^r$ are 2-component Majorana spinors. Since these correspond to broken supersymmetries, we take
\bea
\Gamma_{(4)} \chi &=& -\chi\cr
\Gamma_{(4)} \E_+ &=& \E_+
\eea
Here $\E_+$ is a commuting spinor. We represent the  5d MSYM gamma matrices relevant for us as
\bea
\Gamma_i &=& \gamma_i \otimes 1\cr
\Gamma_{\hat{A}} &=& \gamma_{(4)} \otimes \gamma_{\hat{A}}
\eea
These can be used to construct generators of $SU(2)_+ \times SO(5)$ rotational times R-symmetry.
Here $\gamma_{(4)} = \gamma_{1234}$. We reserve the $5$th index for the M-theory circle. Our R symmetry
indices range over $\hat{A}=6789(10)$. From our realization we see that $\Gamma_{(4)} = \gamma_{(4)} \otimes 1$.
We may then write out all 2-component spinor indices
\bea
\chi_{\alpha}^{\beta'\gamma''} &=& \delta_r A_i q_{i\alpha\dot{\beta}} \E_+^{\dot{\beta}\beta'} \psi^{r\gamma''}
\eea
Here $\beta'$ and $\gamma''$ are 2-component indices such that $\beta'\gamma''$ is a 4-component spinor label of an $SO(5)$ R symmetry spinor, $\psi^{r\gamma''}$ is a 2-component Majorana spinor. Let us first consider the $SU(2)_+$ rotation
\bea
\delta \chi &=& \frac{1}{4}\epsilon^+_{ij} \Gamma_{ij} \chi
\eea
This amounts to
\bea
\delta \chi_{\alpha} &=& \frac{i}{4} \epsilon^+_{ij} \eta_{ij}^{I+} (\sigma^I)_{\alpha}{}^{\beta} \chi_{\beta}
\eea
We next note that
\bea
\frac{1}{4} \epsilon^+_{ij} \eta^{I+}_{ij} (\sigma^I)_{\alpha}{}^{\beta} q_{k\beta\dot\gamma} \E_+^{\dot\gamma} &=& -\epsilon^+_{kj} q_{j\alpha\dot\beta} \E_+^{\dot\beta}
\eea
as a consequence of the gamma matrix identity $\frac{1}{4}\epsilon_{ij}[\gamma_{ij},\gamma_k] = -\epsilon_{kj} \gamma_j$.
This means that we can write
\bea
\delta \chi_{\alpha} &=& -\delta_r A_i \epsilon^+_{ij} q_{j\alpha\dot\beta} \E_{\dot\beta} \psi^{r}
\eea
We next expand
\bea
\epsilon^+_{ij} = \epsilon^{I+} \eta^{I+}_{ij}
\eea
and we use the identity
\bea
\delta_r A_i \eta^{+I}_{ij} &=& -\delta_s A_j (J^{+I})^s{}_r
\eea
where we define
\bea
(J^{I\pm})_{rs} &=& \int \tr \(\delta_r A_i \delta_s A_j\) \eta^{I\pm}_{ij}
\eea
Using the completeness relation of modes \cite{Gauntlett:1993sh} it can be shown that $J^{I\pm}$ obey
the same algebra as $\eta^{I\pm}_{ij}$. We now find the following $SU(2)_+$ action on the Fermi zero modes
\bea
\delta \psi^{r} &=& \epsilon^{I+} (J^{I+})^r{}_s \psi^{s}
\eea

A subgroup of $SO(4)$ R symmetry\footnote{The original $SO(5)$ R symmetry is broken to $SO(4)$ by the vev.} is the $SU(2)_-$ generated by three complex structures. By the same analysis as for $SU(2)_+$ we find that $SU(2)_-$ acts as
\bea
\delta \psi^{r} &=& \epsilon^{I-} (J^{I-})^r{}_s \psi^{s}
\eea

Since $SU(2)_-$ commutes with $SU(2)_+$ we shall associate $J^{I-}$ with $SU(2)_-$. We can also understand the appearance of $J^{I-}$ for the $SU(2)_-$ R symmetry by studying how supersymmetry is induced from 5d MSYM. We have the following supersymmetry variation of the gauge potential,
\bea
\delta A_i &=& i \bar{\omega}_{\dot{\alpha}} q_i^{\dot\alpha\beta} \sigma_1 \chi_{\beta}
\eea
In the moduli approximation we may put
\bea
\delta A_i &=& \delta X^r \delta_r A_i
\eea
by including a gauge variation. We then act by $\int d^4 x\, \tr \, \delta_s A_i$, expand the zero mode $\chi$ in collective coordinates $\psi^r$, and we get
\bea
\delta X^r &=& i \bar{\epsilon} \psi^r + i \bar{\epsilon}^I (J^{I-})^r{}_s \psi^{s}
\eea
Here $\bar{\epsilon} = \bar{\omega} \sigma_1 \E_+$ and $\bar{\epsilon}^I = \bar{\omega} \sigma^I\otimes \sigma_1 \E_+$. By the well-established theory of the $\N=8$ (four complex supercharges) sigma model, we can now reliable identify $J^{I-}$ as the generator of $SU(2)_-$ subgroup of the $SO(4)$ R symmetry generated by the three complex structures on moduli space.

In order to map spinors to forms, we first define complex spinors
\bea
\xi^r &=& \psi^r_1 - i \psi^r_2
\eea
We then map
\bea
\xi^r &\simeq& dX^r
\eea
For the overall $\mb{R}^4$ part of the moduli space on which lives one-forms $dX_i$ ($i=1,...,4$), we further define complex one-forms
\bea
dw_{\dot{1}} &=& dX_4 - i dX_3\cr
dw_{\dot{2}} &=& dX_2 - i dX_1
\eea
The generators are realized as follows. The $SU(2)_+$ generators act on $dX_i$ as
\bea
(J^{I+})_{ij}dX_j &=& -\frac{i}{2}\eta^{I+}_{ij} dX_j
\eea
The $SU(2)_-$ generators act as
\bea
(J^{I-})_{ij}dX_j &=& -\frac{i}{2}\eta^{I-}_{ij} dX_j
\eea
We can establish this by noting that for the overall $\mb{R}^4$ part of the moduli space, the zero modes are
\bea
\delta_j A_i &=& F_{ji}
\eea
From this, we find that $J^{I\pm}_{ij} \sim \eta_{ij}^{I\pm}$. It is important to note that $J^{I-}$ may be identified with the subset $\frac{1}{2}\epsilon_{IJK} R_{JK}$ of the R symmetry generators (\ref{R}). This means that we must have the vector embedding of $SU(2)_- \simeq SO(3)$ into $SO(5)$. In terms of complex coordinates we have
\bea
J^{3+}\(\begin{array}{c}
dw_1\\
dw_2\\
d\bar{w}^1\\
d\bar{w}^2
\end{array}\) = \frac{1}{2}\(\begin{array}{c}
dw_1\\
dw_2\\
-d\bar{w}^1\\
-d\bar{w}^2
\end{array}\), \qquad
J^{3-}\(\begin{array}{c}
dw_1\\
dw_2\\
d\bar{w}^1\\
d\bar{w}^2
\end{array}\) = \frac{1}{2}\(\begin{array}{c}
dw_1\\
-dw_2\\
-d\bar{w}^1\\
d\bar{w}^2
\end{array}\)
\eea
To find the third Cartan generator we change the sign when acting on the last two entries, compared to how $J^{3-}$ acts, so that
\bea
K^3 \(\begin{array}{c}
dw_1\\
dw_2\\
d\bar{w}^1\\
d\bar{w}^2
\end{array}\) &=& \frac{1}{2}\(\begin{array}{c}
dw_1\\
-dw_2\\
d\bar{w}^1\\
-d\bar{w}^2
\end{array}\)
\eea
We can verify that this gives a consistent embedding of $SU(2)_-$ in $SO(4)$ R symmetry by extending this construction to the other generators $J^{I-}$ and construct corresponding generators $K^I$ by again changing the sign when they act on the two last entries. This way we find that $R^{I \pm} = \frac{1}{2}\(J^{I-} \pm K^{I}\)$ generate $SU(2)_L \times SU(2)_R$ which is consistent with the fact that $J^{I-}$ define a vector embedding in $SO(4)$.

Given this, we build a multiplet of states
\bea
\begin{array}{ccccccc}
&&& 1 &&& \\
&& dw_{\dot{\alpha}} &&  d\bar{w}^{\dot{\alpha}} &&\\
& dw_{\dot{1}} dw_{\dot{2}} && dw_{\dot{\alpha}} d\bar{w}^{\dot{\beta}} &&d\bar{w}^{\dot{1}} d\bar{w}^{\dot{2}} &\\
&& d\bar{w}^{\dot{1}} d\bar{w}^{\dot{2}} dw_{\dot{\alpha}} &&  dw_{\dot{1}} dw_{\dot{2}} d\bar{w}^{\dot{\alpha}}&& \\
&&&dw_{\dot{1}} dw_{\dot{2}} d\bar{w}^{\dot{1}} d\bar{w}^{\dot{2}}&&&
\end{array}
\eea
with corresponding weights of Cartan generators $(J^{3+},J^{3-},K^3)$
\bea
\begin{array}{ccccccc}
&&& (0,0,0) &&&
\\
&& \begin{array}{c}
(\frac{1}{2},\pm \frac{1}{2},\pm \frac{1}{2})
\end{array} && \begin{array}{c}
(-\frac{1}{2},\mp\frac{1}{2},\pm \frac{1}{2})
\end{array}&&\\
&(1,0,0) && \begin{array}{c}
(0,0,\pm 1)\\
(0,\pm 1,0)
\end{array} && (-1,0,0)&
\\
&&\begin{array}{c}
(-\frac{1}{2},\pm\frac{1}{2},\pm\frac{1}{2})
\end{array} && \begin{array}{c}
(\frac{1}{2},\mp\frac{1}{2},\pm\frac{1}{2})
\end{array}&&\\
&&&(0,0,0)&&&
\end{array}
\eea
We then recognize that these states fill up a $16$-dimensional massive instanton-particle multiplet
\ben
(3;1,1) \oplus (1;2,2) \oplus (1;1,1) \oplus (2;2,1) \oplus (2;1,2)\label{states16}
\een
of the $SU(2)_+ \times SO(4)$. Here we label the representations of $SO(4)$ by the dimensions of $SO(4) = SU(2)_L \times SU(2)_R$ whose Cartans are
\bea
R^{3\pm} &=& \frac{1}{2} \(J^{3-} \pm K^3\)
\eea
It should be noted that $SU(2)_-$ which is generated by $J^{I-} = R^{I+} + R^{I-}$ is the diagonal subgroup of $SU(2)_L \times SU(2)_R$.

But for the charged instanton this is not the full story as the dyonic instanton-particle also carries internal degrees of freedom whose spin quantum numbers we will obtain in the next section.

\subsection{Relative part of moduli space -- localization to $\mathbb{R}^{4}$ and classification of BPS states}\label{internal}

In this section, we shall 
describe the localization of states to the flat space, $\mathbb{R}^4$, starting from the Eguchi-Hanson space. Since the index is essentially invariant under the scaling of the potential and the corresponding central charge is basically determined by charges, we may compute the 1/4 BPS free energy exactly by taking the limit where the vacuum expectation value (vev) of the scalar field becomes large.  In this limit
the states of the system are  localized around the zeroes (minima) of the potential. We have shown that,
at each localization point, the space becomes  $\mathbb{R}^{4(N-1)}$. At each copy of  $\mathbb{R}^{4}$, the system is described by the ${\cal N}=8$ supersymmetric  quantum mechanics of 4d harmonic oscillators. Below we investigate multiplet structures of this ${\cal N}=8$  quantum mechanics focusing on its BPS sectors.

As we show in appendix \ref{SEguchi}, the Calabi metric when $N=2$, is equivalent with the Eguchi-Hanson metric
\ben
ds^2&=&\alpha^2 \left[ \frac{d\rho^2}{K^2} +\frac{\rho^2}{4}\Big( d\theta^2+ \sin^2\theta d\phi^2+ K^2 (d\psi+\cos \theta d\phi)^2\Big)
\right] \nonumber \\
&=&\alpha^2 \left[ \frac{d\rho^2}{K^2} +\frac{\rho^2}{4}\Big(\sigma_1^2 + \sigma^2_2 + K^2 \sigma_3^2\Big)\right]
\een
by a coordinate transformation. Here $K^2 = 1 - \frac{4\zeta^2}{\rho^4}$, the overall coefficient is $\alpha^2 = 2 \times \frac{4\pi^2}{g^2}$ and
\bea
\sigma_1 + i\sigma_2 &=& e^{i\psi} \(i d\theta + \sin \theta d\phi\)\cr
\sigma_3 &=& d\psi + \cos \theta d\phi\label{sigmaEH}
\eea
where the coordinate ranges are $\phi \in [0, 2\pi]$ and $\psi \in [0, 2\pi]$, such that in the limit $\zeta \rightarrow 0$ the Eguchi-Hanson space degenerates to the orbifold $\mb{C}^2 / \mb{Z}_2$. In particular the Calabi space fiber coordinate $\varphi = 2\phi$ is ranged in $[0,4\pi]$. From the form of the Killing vector in Eq (\ref{G}), which in our case of $N=2$ reduces to
\bea
G &=& v \frac{\partial}{\partial \phi}
\eea
where $v = v_1 - v_2$, we see that the potential takes the form
\be
V= \frac{1}{2} g_{rs}G^r G^s=\frac{1}{2}g_{\phi\phi}v^2=
 \frac{1}{8} \alpha^2 v^2  \left(\rho^2 - \frac{4\zeta^2}{\rho^2}\cos^2 \theta\right)
\ee
The localization occurs at the points where $V$ becomes zero, and there are two localization points for the case of $N=2$:
One is at the north pole $\theta=0$ of the sphere $\rho^2= 2\zeta$ and the other at the south pole $\theta=\pi$
of the same sphere.

Around the north pole, we introduce coordinates
\ben
 \alpha \frac{\rho}{2} K=\tilde\rho
\rightarrow 0  \nonumber\,,  \ \ \
 \alpha \frac{\rho}{2}\,\theta\, =\tilde\theta 
\rightarrow 0
\een
By introducing 
$\tilde\rho=\bar\rho \cos \frac{\bar\theta}{2}$ and $
\tilde\theta = \bar\rho \sin \frac{\bar\theta}{2}$, the metric, to the quadratic order in $\bar\rho$, becomes
\ben
ds^2&=& {d\bar\rho^2} +\frac{\bar\rho^2}{4}\Big[
d \bar\theta^2+ \sin^2 \bar\theta d \psi^2+ (2d\phi+(\cos\bar\theta +1) d\psi)^2\Big]
 \nonumber \\
&=& {d\bar\rho^2}+\frac{\bar\rho^2}{4}\, \Big[ \bar\sigma_1^2 + \bar\sigma^2_2 + \bar\sigma_3^2\Big]
\een
where
\ben
 \bar\sigma_1 + i\bar\sigma_2 &=& e^{i \bar\phi} \Big(
i d \bar\theta + \sin \bar\theta d \bar\psi
\Big)\cr
\bar\sigma_3 &=& d \bar\phi + \cos \bar\theta \, d \bar\psi\label{sigmac}
\een
with
\bea
\bar\phi &=& 2\phi+ \psi\cr
\bar\psi &=& \psi
\eea
Note that $\bar\phi$ is ranged over $[0,4\pi]$. We have two $U(1)$ charges corresponding to two commuting Killing vectors
\bea
q_{\bar{\phi}} &=& -i \L_{\partial_{\bar\phi}}\cr
q_{\bar{\psi}} &=& -i \L_{\partial_{\bar\psi}}
\eea
which can be related to the following two commuting Killing vectors of Eguchi-Hanson metric,
\bea
Q_{\phi} &=& -i \L_{\partial_{\phi}}\cr
Q_{\psi} &=& -i \L_{\partial_{\psi}}
\eea
as
\bea
Q_\psi  &=& q_{\bar{\psi}}+ q_{\bar\phi}\cr
Q_\phi &=& 2 q_{\bar\phi}
\eea
Here $Q_\psi$/$q_{\bar{\psi}}$ and $Q_\phi$/$q_{\bar{\phi}}$ are integral/half-integral quantized.

Around the south pole, we introduce coordinates
\ben
 \alpha \frac{\rho}{2} K=\tilde\rho
\rightarrow 0\,, \ \ 
 \alpha \frac{\rho}{2}\,(\pi-\theta)\, =\tilde\theta 
\rightarrow 0
\een
By introducing 
$\tilde\rho=\bar\rho \sin \frac{\bar\theta}{2}$ and $
\tilde\theta = \bar\rho \cos \frac{\bar\theta}{2}$, the metric, to the quadratic order in $\bar\rho$, becomes
\ben
ds^2&=& {d\bar\rho^2} +\frac{\bar\rho^2}{4}\Big[
d \bar\theta^2+ \sin^2 \bar\theta d \psi^2+ (2d\phi+(\cos\bar\theta -1) d\psi)^2\Big]
 \nonumber \\
&=& {d\bar\rho^2}+\frac{\bar\rho^2}{4}\, \Big[ \bar\sigma_1^2 + \bar\sigma^2_2 + \bar\sigma_3^2\Big]
\een
where
\bea
\bar\phi &=& 2\phi- \psi\cr
\bar\psi &=& -\psi
\eea
The angle $\bar\phi$ is again ranged over $[0,4\pi]$ and
\bea
Q_{-\psi} &=& q_{\bar{\psi}}+ q_{\bar\phi}\cr
Q_\phi &=& 2 q_{\bar\phi}
\eea
where the minus sign in front of $\psi$ reflects the change of the relative orientation of the tangent space at the south pole
in comparison with that of the north pole. We furthermore have that $Q_{-\psi} = - Q_{\psi}$.
Introducing vielbeins by
\be
\bar{e}^0 = d \bar\rho,\ \ \ \ \bar{e}^I =\frac{\bar{\rho}}{2} \bar\sigma_I
\ee
the metric for $\mathbb{R}^4$ takes the form
\be
ds^2= \bar{e}^0 \bar{e}^0 +\bar{e}^I\bar{e}^I
\ee
Furthermore,
\bea
G &=& \bar{v} \partial_{\bar\phi},\cr
\bar{v} &:=& 2v
\eea
and the potential becomes
\bea
V &=& \frac{1}{8} \bar{v}^2 \bar{\rho}^2
\eea
locally near the north or the south pole.

In order to construct the generators for the $SU(2)_L$ part of the R-symmetry, we need the expressions for the three complex structures
given by
\be
I_I = e^0 i_{e^I}-e^I i_{e^0} +\epsilon_{IJK} e^J i_{e^K}
\ee
Upon localization, they are reduced to
\be
I_I = E^0 i_{E^I}-E^I i_{E^0} +\epsilon_{IJK} E^J i_{E^K}
\label{complex}
\ee
where the new set of vielbein is defined as
\ben
 E^1 + i E^2 &=& \frac{\bar\rho}{2}e^{i \bar\psi} \Big(
i d \bar\theta + \sin \bar\theta d \bar\phi
\Big), \nonumber\\
  E^0+iE^3 &=&d \bar\rho+i\frac{\bar\rho}{2} (d \bar\psi + \cos \bar\theta \, d \bar\phi)
\een
These satisfy
\be
I_I I_J = -\delta_{IJ} + \epsilon_{IJK} I_K
\ee

\subsubsection{BPS states in $\mathbb{R}^4$}
In this subsection, we would like to describe the general structure of  BPS states of the ${\cal N}=8$ supersymmetric harmonic oscillator in $\mathbb{R}^4$. We are interested in the solutions of the BPS equation
\ben
\Big[(d- i_G)+i (d^\dagger -G)\Big]\Omega=0\label{BPSEQ}
\een
where $ d^\dagger = - * d *$.
Since the BPS operator ${\cal Q}-i {\cal Q}^\dagger$ is commuting with the self-dual or anti-self-dual projections, one can separate states into a sum of self-dual and anti-self-dual parts
\bea
\Omega = \Omega_+ + \Omega_-
\eea
Within the BPS sector, the even-form and the odd-form part of the wave functions are decoupled from each other. We shall call the even-form/odd-form part as bosonic/fermionic, the meaning of which will be clear when we discuss the spin content of BPS multiplets. The BPS states are characterized by the central charge
\be
{Z}= \bar{v} \, q \, \ge 0
\ee
where the charge $q$ is half-integral quantized
with the charge 
operator
\be
\hat{q}= -i {\cal L}_{\partial_{\bar\phi}}
\ee
Below we shall concentrate on the BPS states with $\bar{v}>0$ and $q\ge 0$ and the case with  $\bar{v}< 0$ and $q\le 0$ will be briefly discussed at the end. The BPS solutions exist only in the self-dual sector. To classify the solutions we will use spherical coordinates $(\rho,\bar{\theta},\bar{\phi},\bar{\psi})$ on $\mb{R}^4$ and we introduce the Wigner D-function
\bea
D^j_{mq} &=& \<jm|\sigma|jq\>
\eea
where $\|jm\>$ denotes a spin-$j$ state of $SU(2)$ with $m,q = -j,-j+1,\cdots,j$. We define
\bea
\bar{\sigma} &=& \bar{g}^{-1}d\bar{g}\cr
\bar{g} &=& e^{i \bar{\psi} J_3} e^{i \bar{\theta} J_2} e^{i \bar{\phi} J_3}
\eea
where $J_I$ generate $SU(2)$ with commutation relations $[J_I,J_J] = i\epsilon_{IJK} J_K$. The components of $\bar{\sigma} = \bar{\sigma}_I J_I$ are given in (\ref{sigmac}). Expressing the D-function as
\bea
D^j_{mq} &=& e^{im \bar{\psi}+iq\bar{\phi}} d^j_{mq}(\bar{\theta})
\eea
we see that
\bea
\hat{q} D^j_{mq} &=& q D^j_{mq}
\eea
When $q=0,1/2,1,3/2,\cdots$ we find the following bosonic multiplet of charge $\hat{q} = q$ states as solutions to the 1/4-BPS equation (\ref{BPSEQ})
\ben
\Omega^q_{mq} &=& D^q_{mq}\, \bar{\rho}^{2q} \, e^{-\frac{\bar{v}}{4} \bar\rho^2} (1+ \bar{e}^1\bar{e}^2)(1+ \bar{e}^0\bar{e}^3) \\
\Omega^{q-1}_{mq} &=& D^{q-1}_{m, q-1}\, \bar{\rho}^{2(q-1)} \, e^{-\frac{\bar{v}}{4} \bar\rho^2}( \bar{e}^0+i\bar{e}^3 ) (\bar{e}^1+i \bar{e}^2)
\een
Our notation is such that the states $\Omega^j_{mq}$ carry charge $\hat{q} = q$ and fall in a $j$-multiplet of
states with $m=-j,-j+1,\cdots,j$. We will occasionally suppress the labels $mq$ and write these
states as $\Omega^q$ and $\Omega^{q-1}$. The $j=q-1$ multiplet exists only for $q \ge 1$ and has
charge $\hat{q}=q$ as a consequence of $\hat{q} (\bar{e}^1+i\bar{e}^2) = 1$. We obtain these
states in appendix \ref{Instanton} by solving the BPS equation. By acting with the
supercharge ${\cal Q}+i {\cal Q}^\dagger$ on these bosonic states, we obtain
two $j= q-\frac{1}{2}$ multiplets $\Omega^{\pm,q-\frac{1}{2}}$ when $q \ge \frac{1}{2}$.
These multiplets are fermionic and we will describe their spin content below. Therefore, for $q>0$, we have the following multiplet of 1/4 BPS states
\ben
(2q+1)\  \oplus\  (2q-1)\  \oplus \  2q^+ \ \oplus \ 2q^-\label{states}
\een
Here representations of $SU(2)$ are labeled by their dimension. The physical role of this $SU(2)$ will be clarified shortly and will be identified as the unbroken $SU(2)_+$ Lorentz symmetry. Thus we find in total $8q$ states for $q=1/2,1,3/2,\cdots$. For $q=0$ we have on the other hand a unique state
\be
\Omega_0= e^{-\frac{\bar{v}}{4} \bar\rho^2}  (1+ \bar{e}^1\bar{e}^2)(1+ \bar{e}^0\bar{e}^3)
\ee
which is annihilated by all supercharges
\be
{\cal Q} \, \Omega_0 = {\cal Q}^\dagger \, \Omega_0=0
\ee
Therefore no new odd-form state is generated by the action of the supercharges and the corresponding $q=0$ state is unique.

For $\bar{v}$ negative, the charge $q$ has to be non-positive definite, since the central charge ${\cal Z}$ is non-negative definite, and
the corresponding
number of degenerate states remains. To show  this, one may use the parity symmetry
of the system
\be
\bar{\psi} \rightarrow  \bar\psi +\pi, \ \bar{\theta} \rightarrow \pi-\bar\theta,\ \bar{\phi} \rightarrow -\bar\phi,
\ee
and
\be
\bar{v} \rightarrow  -\bar{v}
\ee
Consequently, one finds the same number of states as is demonstrated in the appendix explicitly.
Therefore, the number of degenerate BPS states with charge $q$ is
\bea
n_q &=& \left\{
\begin{array}{cl}
8 |q|=4 |Q| &   {\mbox{if}}\ \bar{v}q > 0  \\
1     &   {\mbox{if}}\  q=0 \ {\mbox{and}}\ \bar{v} \neq 0\\
0    &    {\mbox{if}}\ \bar{v}q < 0
\end{array}
\right.
\eea
where $Q$ is the eigenvalue of the integral-valued (F-string) charge
\be
\hat{Q}=2 \hat{q}= -i {\cal L}_{\partial_\phi}
\ee

\subsubsection{Rotation- and R-symmetries}

The Eguchi-Hanson space is invariant under the action of $SU(2)/\mathbb{Z}_2 \times U(1)$ transformation. The  $SU(2)/\mathbb{Z}_2$
isometry generated by the Killing vectors
\ben
L^{EH}_1 + i L^{EH}_2 &=& e^{i \phi} \Big( i\partial_\theta- \cot \theta \partial_\phi +\frac{1}{\sin \theta} \partial_\psi \Big)
\nonumber
\\
 L^{EH}_3 &=& \partial_\phi
\een
acts triholomorphically as
\be
{\cal L}_{L^{EH}_I} I^J =0
\ee
as a consequence of $\L_{L_I^{EH}} \sigma_J = 0$ where $\sigma_J$ are the Maurer-Cartan forms given by
Eq.~(\ref{sigmaEH}). The remaining $U(1)$ generated by  $\partial_\psi$
has nothing to do with the F-string charge.  As derived in the context of ADHM construction,  the triholomorphic vector field $G$ relevant for the F-string charge
has to do with  the other  $U(1)$ 
generated by $\partial_\phi$. In the presence of the potential, the $SU(2)/\mathbb{Z}_2$ symmetry of the quantum mechanical system is further broken down to the $U(1)_\phi$. Hence once the vev of the scalar is turned on, the
quantum mechanics is no longer invariant under  $SU(2)/\mathbb{Z}_2 \times U(1)_\psi$\footnote{Note here that, without introducing
the noncommutativity,
the $U(1)_\psi$ is enhanced to $SU(2)_b$ symmetry, which apparently rotates the noncommutativity parameter $\theta_{ij}$.}  but only
under  $U(1)_\phi\times U(1)_\psi$. The $SO(4)$ little group of the $SO(5)$ R-symmetry will remain and below we shall focus on the $SO(3)$ subgroup
of this $SO(4)$. As discussed in the previous section, this $SO(3)$ R symmetry  is generated by
the action of the three complex structures $I^I$. Their action on a form is multiplicative and satisfies the usual Leibniz rule:
\ben
I_I dx^i = dx^j (I_I)_j \,^i
\een
and
\be
I_I dx^i\wedge dx^j = (I_I dx^i ) \wedge dx^j + dx^i \wedge (I_I dx^j )
\ee

We would now like to understand the multiplet structure of the states we have constructed
by the localization to $\mathbb{R}^4$.  In case of the $R$ symmetry, the story is rather clear since the complex structures $I^I$ of the Eguchi-Hanson space has a natural realization in $\mathbb{R}^4$:  Their explicit form   in $\mathbb{R}^4$
is given in (\ref{complex}). Next we would like to identify the $SU(2)_+$ spatial rotation. Since $\mathbb{R}^4$ has $SU(2) \times SU(2)$ symmetry, it is rather clear that a particular combination of the $SU(2)$'s realizes the $SU(2)_+$. Let us denote the two $SU(2)$'s by
$SU(2)_\phi$ and $SU(2)_\psi$ where $U(1)_\phi$ and $U(1)_\psi$ subgroups are included into
 $SU(2)_\phi$  and $SU(2)_\psi$  respectively.

For reasons described below, we find that $SU(2)_+ = SU(2)_\psi$.
First of all, the multiplets are labeled by a fixed $U(1)_\phi$ charge, which is  physically interpreted as F-string  charge
of the dyonic instantons. If the $SU(2)_+$ rotation involves this $U(1)_\phi$, then there is no way to
understand the multiplet structure of the above states.
Secondly the $U(1)_\phi$ belongs to $SU(2)/\mathbb{Z}_2$ of the EH space which is originated from the $SU(N)$ gauge symmetry
of the SYM theory instead of any spacetime symmetries. Finally, with the choice of $SU(2)_\psi$ for the rotation symmetry,
one can understand the full multiplet structure in a natural manner as we shall demonstrate shortly. A similar choice
can be realized for the general $SU(N)$ Calabi space upon localization and the rotational $SU(2)_+$ in each $\mathbb{R}^4$
should be chosen such that this $SU(2)_+$  do not include the $U(1)$ responsible for the F-string charges.

By localization the symmetry $U(1)_\psi$ is enhanced to $SU(2)_{\psi}$. On the other hand, by turning off the noncommutativity parameter the rotation symmetry $SU(2)_+$ gets enhanced to $SU(2)_+ \times SU(2)_b$. Very naively then, since we have enhancements to $SU(2)_{\psi}$ and $SU(2)_b$ both being related to the noncommutativity parameter in various limits, one may think that these two $SU(2)$ shall be identified. But this is incorrect. Localization limit effectively means sending noncommutativity parameter to infinity which is a completely different limit from turning off the noncommutativity parameter. The noncommutativity parameter is related to the localization point because the potential, and its zeroes, depend on the noncommutativity parameter. As we approach the localization point the $U(1)_{\psi}$ symmetry is enhanced to $SU(2)_{\psi}$ and the localization point is not rotated by $SU(2)_{\psi}$ which is a tangent space symmetry at the localization point. Neither is the noncommutativity parameter rotated by $SU(2)_+$. This shows that we cannot identify $SU(2)_\psi$ with $SU(2)_b$ which would rotate the noncommutativity parameter. We conclude that we must identify $SU(2)_{\psi} = SU(2)_+$ since $SU(2)_+$ leaves the noncommutativity parameter fixed, just like the tangent space group $SU(2)_{\psi}$ at the localization point is supposed to leave the localization point fixed.

The $SU(2)_\psi$ is generated by the
the Killing vectors
\ben
 \bar{L}_1 + i \bar{L}_2 &=& e^{i \bar\psi} \Big( i\partial_{\bar\theta} - \cot \bar\theta \partial_{\bar\psi}
+\frac{1}{\sin \bar\theta} \partial_{\bar\phi} \Big)
\nonumber
\\
 \bar{L}_3 &=& \partial_{\bar\psi}
\een
The orbital angular momentum is generated by the Lie derivatives
\be
M_I = -i {\cal L}_{\bar{L}_I}= -i (d i_{\bar{L}_I}+ i_{\bar{L}_I} d )
\ee
It is straightforward to demonstrate that the vielbein $\bar{e}^a=(\bar{e}^0, \bar{e}^I)$ is invariant under the action of
$M_I$:
\be
M_I \bar{e}^a=0
\ee
One then recognizes that the states $\Omega^q$ and $\Omega^{q-1}$ fall in $(2q+1)$ and $(2q-1)$ dimensional representations of $SU(2)_{\psi}$.

The R-symmetry generators
\be
R_I = \frac{i}{2} I_I
\ee
satisfy the $SU(2)$ algebra
\be
[R_I, R_J]= i\epsilon_{IJK}R_K
\ee
It is  straightforward to show that $R_I$ transforms
\be
[M_I, R_J]= i \epsilon_{IJK}R_K
\ee
as a triplet
under $M_I$.  The desired total angular momentum including the spin part
should be chosen as \cite{Bak:1999ip}
\be
J_I = M_I - R_I
\ee
which satisfy the $SU(2)$ algebra
\be
[J_I, J_J]= i\epsilon_{IJK}J_K
\ee
Furthermore,  one finds that the R-symmetry and the rotation generator commute, i.e.
\be
[J_I, R_J]=0
\ee
which is required from the first principle construction of the generators starting from the SYM theory.
One finds that $\Omega^q$ and $\Omega^{q-1}$ are $SU(2)_L$ R-symmetry singlets,
\ben
R_I \Omega^q &=& 0\cr
R_I \Omega^{q-1} &=& 0
\een
so that
\be
\Omega^q = (2q+1,1)\,, \ \  \Omega^{q-1}=(2q-1,1)
\ee
where the first  and the second numbers in the bracket denote the dimensions of  representations of $SU(2)_+ \times SU(2)_L$.

Let us now turn to the case of odd-forms. As we said before, the odd-forms can be generated by
applying the combination of supercharges to the even-form solutions. To understand the corresponding multiplet structure,
we note that  the supercharges transform as a singlet plus a triplet under $M_I=J_I+R_I$,
\be
[M_I, {\cal Q}]=0\, \ \ [M_I,{\cal Q}_J]=i\epsilon_{IJK}{\cal Q}_K
\ee
As the four supercharges belong to doublets under $R_I$, they must form  doublets under $J_I$ as well.  Indeed one may construct one doublet of $J_I$ by
\be
{\cal Q}^-_+ = i ({\cal Q}^3 -i {\cal Q}^4 )\,, \ \ {\cal Q}^-_-=i({\cal Q}^1-i {\cal Q}^2)
\ee
with $[R_3, {\cal Q}^-_\pm]= -\frac{1}{2} {\cal Q}^-_\pm$. The second combination
\be
{\cal Q}^+_+=i({\cal Q}^1+i {\cal Q}^2)\,, \ \ {\cal Q}^+_- = -i ({\cal Q}^3+i {\cal Q}^4 )
\ee
forms a doublet under $J_I$ with $[R_3, {\cal Q}^+_\pm]= \frac{1}{2} {\cal Q}^+_\pm$.
Then by the action of the appropriate combination of ${\cal Q}^-_\pm$ to $\Omega^q\oplus \Omega^{q-1}$, one generates states $\Omega^{-,q-\frac{1}{2}}$ while, by ${\cal Q}^+_\pm$, one generates states $\Omega^{+,q-\frac{1}{2}}$. Thus the odd form states form the representation
\ben
\Omega^{+,q-\frac{1}{2}} \oplus \Omega^{-,q-\frac{1}{2}} &=& (2q,2)
\een
For $q=0$, there is a unique state
\ben
\Omega_0 &=& (1,1)
\een
which is 1/2 BPS. The minimal $q=\frac{1}{2}$ multiplet is
\ben
(\Omega^{\frac{1}{2}}) \oplus (\Omega^{+,0} \oplus \Omega^{-,0}) &=& (2,1)\oplus (1,2)
\een
which consists of $4$ states.

The 1/4-BPS dyonic instanton multiplet with $64$ states \cite{Lambert:2010iw} can be obtained by taking the tensor product of the
 $16$ states in $\mb{R}^4$ with the above $4$ states at the localization point,
\bea
&& \Big((3;1,1) \oplus (1;2,2) \oplus (1;1,1) \oplus (2;2,1) \oplus (2;1,2)\Big) \cr
& \otimes & \Big((2;1,1)\oplus (1;2,1)\Big)
\eea
Here we have included the trivial representation (which is $1$) of the additional representation of an $SU(2)_R$
which is inside the full unbroken $SO(4) = SU(2)_L \times SU(2)_R$ R-symmetry and which is not generated by the three Kahler forms which generate the $SU(2)_L$. Thus the above denotes representations of $SU(2)_+ \times (SU(2)_L\times SU(2)_R)$. Expanding
out the tensor product we recover the multiplet of \cite{Lambert:2010iw}.

\section{Refined partition function and Index}\label{refined}
The spin content can be seen by computing a refined partition function
\bea
Z(\beta,a,b,c) &=& \tr_{1/4\,{\rm BPS}} \(e^{-\beta H} e^{J^{3+} a + R^{3+} b + R^{3-} c}\)
\eea
where $J^{3+},R^{3\pm}$ are Cartans of $SU(2)_+\times SU(2)_L\times SU(2)_R$. For BPS states over which we trace, the Hamiltonian can be replaced by the central charge. For $N=2$ this is given by $Z= vQ = 2vq$. Furthermore, as we have identified $SU(2)_+ = SU(2)_{\psi}$, we shall take
\bea
J^{3+} = \pm Q_{\psi} =  q + q_{\bar\psi}
\eea
on north and south pole respectively. Using this, we get
\bea
Z &=& \tr_{1/4\,{\rm BPS}} \(e^{- q \(2\beta v - a\)} e^{q_{\bar{\psi}} a + R^{3+} b + R^{3-} c}\)
\eea
and explicitly
\bea
Z &=& 1 + \sum_q e^{-q \(2\beta v - a\)} \[s(2q+1,1) + s(2q-1,1) + s(2q,2)\]
\eea
where we define
\bea
s(2j+1,2k+1) &=& \sum_{m,n} e^{am+bn}
\eea
and $m=-j,-j+1,\cdots, j-1,j$ and $n=-k,-k+1,\cdots, k-1,k$. We find
\bea
s(2j+1,2k+1) &=& \frac{\sinh\(\frac{a}{2}(2j+1)\)\sinh\(\frac{b}{2}(2k+1)\)}{\sinh\frac{a}{2}\sinh\frac{b}{2}}
\eea
We get
\bea
Z &=& \frac{\cosh \frac{2\beta v-a-b}{4} \cosh \frac{2\beta v-a+b}{4}}{\sinh \frac{\beta v-a}{2} \sinh \frac{\beta v}{2}}
\eea
We can now also compute the index
\bea
{\mbox{Index}} &=& \tr\((-1)^F e^{-\beta H} e^{J^{3+} a + R^{3+} b + R^{3-} c}\)
\eea
as follows
\bea
{\mbox{Index}} &=& 1 + \sum_q e^{-q \(2\beta v - a\)} \[s(2q+1,1) + s(2q-1,1) - s(2q,2)\]
\eea
with the result
\bea
{\mbox{Index}} &=& \frac{\sinh \frac{2\beta v-a-b}{4} \sinh \frac{2\beta v-a+b}{4}}{\sinh \frac{\beta v-a}{2} \sinh \frac{\beta v}{2}}
\eea
We may notice that
\bea
{\mbox{Index}}(b+2\pi i) &=& Z(b)
\eea
and indeed this relation can be explained by noticing that $e^{2\pi i R^{3+}} = (-1)^F$. Finally we notice that our result agrees with \cite{Kim:2011mv} if we make the following identifications
\bea
a &=& 2i \gamma_R\cr
b &=& 2i \gamma_2\cr
\beta v &=& \mu
\eea
where definitions of parameters on the right-hand side are found in \cite{Kim:2011mv}.

In the index we do not need to restrict ourselves to 1/4-BPS states since all non-BPS states are paired by a superpartner state with opposite $(-1)^F$. But for the partition function over 1/4-BPS states we can not drop the projection onto 1/4-BPS states. However this can again be expressed as an index by noting that
\bea
\tr_{\mbox{non-BPS}} (-1)^{F+2R^{3+}} &=& 0
\eea
To see this we first note that none of the supercharges $\Q_{\pm}^{\pm}$ can annihilate a non-BPS state. As we act with a sequence of these supercharges on some bosonic/fermionic non-BPS state they will
generate $8$ states with $(-1)^F = +1/-1$ representations
$(3;1,1)\oplus (1;1,1) \oplus (1;2,2)$ and $(-1)^F = -1/+1$ representations $(2;2,1)\oplus (2;1,2)$ (tensor multiplied with the representation of the non-BPS states with which we started). By inspection we see that the sum of $(-1)^{F+2R^{3+}}$ cancels for the 8 bosonic and 8 fermionic states separately. We can now express the 1/4-BPS partition function as the following index
\bea
Z(\beta,a,b,c) &=& \tr \((-1)^{F+2R^{3+}} e^{-\beta H} e^{J^{3+} a + R^{3+} b + R^{3-} c}\)
\eea
where we can drop the explicit projection onto 1/4-BPS states due to the cancelation between non-BPS states as we argued for above.

\section{The commutative limit}
Our index and partition function do not depend on the noncommutativity parameter $\zeta$. We claim that noncommutativity parameter can be smoothly taken towards zero. We can justify this claim for U(2) gauge group. In the commuting limit the Eguchi-Hanson space becomes the orbifold $\mb{C}^2/\mb{Z}_2$ with metric
\bea
ds^2 &=& d\rho^2 + \frac{\rho^2}{4} \(d\theta^2 + \sin^2 \theta d \phi^2 + (d\psi + \cos \theta d\phi)^2\)\cr
&=& d\rho^2 + \frac{\rho^2}{4} \(d\theta^2 + \sin^2 \theta d\psi^2 + (d\phi + \cos \theta d\psi)^2\)
\eea
where $\phi$ and $\psi$ are $2\pi$ ranged. The fact that we can exchange $\psi$ and $\phi$ in this metric can be traced to the fact that $S^3$ can be described either in terms of left-invariant or right-invariant Maurer-Cartan forms. This symmetry is present only in the commutative limit and is not a symmetry of the Eguchi-Hanson metric.

By now substituting $\theta,\psi,\phi$ with $\bar{\theta},\bar{\psi},\bar{\phi}$, we see that we have already obtained all these
1/4-BPS solutions in Eq.~(\ref{states}). The only difference is that here $\phi$ is $2\pi$-ranged instead of $4\pi$-ranged, so that the corresponding electric charge $Q_{\phi} = - i \L_{\partial_\phi}$ is integer quantized. Let us denote the charge integer-eigenvalue on a state by $Q$. Then the bosonic states with charge $Q$ are given by
\bea
\Omega^Q_{mQ}|_{\zeta=0} &=& D^Q_{mQ} \rho^{2q} e^{-\frac{v\rho^2}{4}} \(1+e^1 e^2\)\(1+e^0 e^3\)\cr
\Omega^{Q-1}_{mQ}|_{\zeta=0} &=& D^{Q-1}_{m,Q-1} \rho^{2(Q-1)} e^{-\frac{v\rho^2}{4}}\(e^0+ie^3\)\(e^1+ie^2\)
\eea
Let us comment that it seems out of reach to find corresponding exact BPS solutions away from $\zeta = 0$ where we instead must rely on localization computations.

These states carry $U(1)$ charges that is most conveniently labeled by $Q_{\phi} = Q$ and by $Q_{\psi} = m$. To understand that the commutative limit is smooth, we need to match the $U(1)$ charges of these states with corresponding $U(1)$ charges of the states we found on the Eguchi-Hanson space by the localization computation. Since $Q$ is integer quantized and $q$ from localization computation are half-integer quantized, we must have that
\bea
Q &=& 2q
\eea
since otherwise we could never hope to match these states in a one-to-one fashion. On the north pole we have the states $\Omega^q_{mq}$ and $\Omega^{q-1}_{mq}$ and we have corresponding states on the south pole. All these states carry charge $Q_{\phi}=2q$. We then recall the relations
\bea
Q_{\psi} &=& \pm\(q_{\bar{\psi}} + q_{\bar{\phi}}\)
\eea
where $+$ is for the north pole and $-$ is for the south pole. Then we find on the north pole that the $j=q$ multiplet has $Q_{\psi} = m + q = 0,1,\cdots,2q$ and the $j=q-1$ multiplet has $Q_{\psi} = m + q = 1,2,\cdots,2q-1$. On the south pole we find that the $j=q$ multiplet has $Q_{\psi} = -2q,\cdots,0$ and the $j=q-1$ multiplet has $Q_{\psi} = -(2q-1),\cdots,-1$. Thus collecting the states, and recalling that $Q=2q$, we see that we indeed have states with
\bea
Q_{\psi} &=& -Q,\cdots,Q\cr
Q_{\psi} &=& -(Q-1),\cdots,Q-1
\eea
which matches with the multiplet of states $\Omega^Q|_{\zeta=0} \oplus \Omega^{Q-1}|_{\zeta=0}$ that we found above.
The odd-form parts can be related in a similar manner.

Let us finally confirm that these bosonic states in the orbifold limit are really singlets under the $SU(2)_L$ R symmetry generated by the three complex structures on the Eguchi-Hanson space in the orbifold limit. We have the Kahler form
\bea
K_3 &=& e^0 e^3 + e^1 e^2
\eea
and indeed the corresponding complex structure
\bea
I_3 &=& e^0 i_{e^3} - e^3 i_{e^0} + e^1 i_{e^2} - e^2 i_{e^1}
\eea
leaves all the bosonic states invariant,
\bea
I_3 \Omega^Q|_{\zeta=0} &=& 0\cr
I_3 \Omega^{Q-1}|_{\zeta=0} &=& 0
\eea
We thus find exactly the same states in the orbifold limit and so we conclude that the
orbifold limit appears to be smooth, although we do not have a direct proof for this for the higher-dimensional Calabi spaces.

\section{Discussion}
We have computed the index and the partition function of one 1/4-BPS dyonic instanton
in noncompact 5d MSYM with U(N) gauge group being maximally broken to U(1)$^{N-1}$ by a generic vev of one
of the five scalar fields, which induces a potential term in the corresponding sigma model. The number of states does not quite
sum up nicely to the anomaly coefficient $\sim N(N^2-1)$ but probably there is no reason to expect this number to emerge
here as we only consider the $k=1$ instanton sector.

One obvious direction to look at further is the higher $k$ generalization. The general form of the metric and the potential  are not known yet
especially with the noncommutativity turned on.  We leave this for the future study.

 It would be very interesting if one can understand  what happens when the
gauge group is not maximally broken.

We can also ask what happens if we compactify one direction
of D4 on a circle. In that case we expect the theory to have an S-duality, and it would be interesting to confirm that the 1/4-BPS
dyonic instanton states and the monopole-string states carry the same spin quantum numbers so that they can be mapped
into each other under S-duality \cite{Tachikawa:2011ch,Kim:2011mv}. As it was argued in \cite{Lambert:2012qy},
showing that 5d MSYM is S-dual would also give strong evidence that 5d MSYM and the corresponding 6d $(2,0)$ theory on a circle, are equivalent.

\subsection*{Acknowledgements}
DB would like to thank Hee-Cheol Kim,  Kimyeong Lee and  Soo-Jong Rey  for helpful discussions. This work was
supported in part by NRF SRC-CQUeST-2005-0049409 and  NRF Mid-career Researcher
Program 2011-0013228.

\appendix

\section{Spinor conventions}\label{Spinor}
We represent the 11d gamma matrices as
\bea
\Gamma_0 &=& \gamma_{(4)} \otimes i\sigma_2 \otimes 1\cr
\Gamma_i &=& \gamma_i \otimes 1 \otimes 1\cr
\Gamma_5 &=& \gamma_{(4)} \otimes \sigma_1 \otimes 1\cr
\Gamma_{\hat{A}} &=& \gamma_{(4)} \otimes \sigma_3 \otimes \gamma_{\hat{A}}
\eea
where we split $\mu = (0,i)$ and $i=1,2,3,4$. We define $\gamma_{(4)} = \gamma_{1234}$. We let ${\hat{A}}=6,7,8,9,(10)$ and reserve the index $5$ for the M-theory circle. The 11d charge conjugation matrix is chosen as
\bea
C &=& \Gamma_0
\eea
We represent $SO(4)$ gamma matrices in quaternion Weyl basis
\bea
\gamma_i &=& \(\begin{array}{cc}
0 & q_i^{\dot{\alpha} \beta}\\
\bar{q}_{i\alpha\dot{\beta}} & 0
\end{array}\)
\eea
Here
\ben
q_i &=& (-i\sigma_I,1)\cr
\bar{q}_i &=& (i\sigma_I,1)\label{quaternions}
\een
are quaternions and their conjugates in the $2\times 2$ representation where
\bea
\sigma_1 = \(\begin{array}{cc}
0 & 1\\
1 & 0
\end{array}\),\quad \sigma_2 = \(\begin{array}{cc}
0 & -i\\
i & 0
\end{array}\), \quad \sigma_3 = \(\begin{array}{cc}
1 & 0\\
0 & -1
\end{array}\)
\eea
are the Pauli matrices. We have the relations
\bea
q_i \bar{q}_j &=& \delta_{ij} +i \eta_{ij}^{I-}\sigma_I\cr
\bar{q}_i q_j &=& \delta_{ij} +i \eta_{ij}^{I+}\sigma_I
\eea
where the selfdual and antiselfdual 't Hooft tensors are given by
\bea
\eta_{ij}^{I\pm} &=& \epsilon_{Iij4} \pm
(\delta^I_i\delta^4_j-\delta^I_j\delta^4_i)
\eea

\section{Flat metric on $\mathbb{H}= \mathbb{R}^4$}\label{Flat}
Let us begin with a description of  $\mathbb{H}$. We introduce a coordinates $y=y_i q_i$ in $\mb{H}$ where $(y_i)\in \mb{R}^4$. Thus $y_i \mapsto y = y_i q_i$ is a map $\mb{R}^4 \rightarrow \mb{H}$. The flat metric reads
\be
ds^2= dyd\bar{y}= dy_i dy_i
\ee
Note that  the quarternion $y$ can be represented as
$y= a e^{q_3 \psi/2}$ with $a$ being purely imaginary, i.e. $a= -\bar{a}$ .
We further introduce
\be
4{x_I}q_I=y q_3 \bar{y}=aq_3 \bar{a}
\ee
With this definition,  one finds
\ben
&& 4x_3 = y^2_3+y^2_4-(y_1^2 +y_2^2)= a^2_3-(a_1^2 +a_2^2)\\
&& 2(x_1 + i x_2) =  y_3 y_1 + y_4 y_2+i (y_3 y_2-y_4 y_1)=a_3( a_1 +i a_2)
\een
Representing $a$ by
\be
a= 2\sqrt{x} \sin \frac{\theta}{2}\Big(q_1\cos \phi+q_2 \sin \phi)+2 q_3 \sqrt{x} \cos \frac{\theta}{2}
\ee
one finds
\be
x_I q_I = x\Big(\sin \theta\, (q_1\cos \phi+q_2 \sin \phi)+ q_3  \cos {\theta}
\Big)
\ee
The flat metric $dqd\bar{q}$ then becomes
\be
ds^2= dad\bar{a}+ \frac{1}{4}a\bar{a} d\psi^2 +\frac{1}{2} d\psi(a q_3 d\bar{a}-da q_3\bar{a})
\ee
Introducing $b$ by $a=2 \sqrt{x}b$, the metric can be presented as
\be
ds^2 = \frac{dx_adx_a}{x}+
{x} \, (d\psi + bq_3 d \bar{b}-db q_3  \bar{b})^2
\ee
Note that
\be
d(b q_3 d \bar{b}-db q_3  \bar{b})= *_3 d \frac{1}{x}
\ee
By explicit compution, one can show
\be
\sigma_\psi = d\psi + (bq_3 d \bar{b}-db q_3  \bar{b})
=d\psi +A=
 d\psi +(\cos \theta -1) d\phi
\ee
Therefore one is led to
\be
ds^2= 
\frac{d\vec{x}^2}{x} + x \sigma^2_\psi
\ee

\section{Calabi metric from the caloron dynamics}\label{Calabi}
In this section we shall derive the Calabi metric for the $k=1$ instanton starting from the known caloron dynamics. This will
be helpful in understanding the corresponding brane picture. We begin with the metric for the $U(N)$ caloron \cite{Lee:1999xb},
\be
ds^2= \frac{4\pi^2}{g^2} L_4 \left[ \,
M_{uv} d\vec{y}_u \cdot d\vec{y}_v +M^{-1}_{uv} \sigma_{\xi_u} \sigma_{\xi_v}
\right]
\ee
where
\be
M_{uv}d\vec{y}_u \cdot d\vec{y}_v =
\sum^N_{u=1} m_u \, d \vec{y}^2_u + \frac{d \vec{y}^2_{1N}}{y_{1N}}+\frac{d \vec{y}^2_{21}}{y_{21}}\cdots
+\frac{d \vec{y}^2_{NN-1}}{y_{NN-1}}
\ee
with
\ben
 \vec{y}_{uv}= \vec{y}_u -\vec{y}_v + \frac{ \zeta}{L_4}\, \delta_{u}^{N} \delta_{v}^{ N-1}
\een
The coordinate $\xi_u$ is ranged over $[0, 4\pi]$ and we introduce
\be
 \sigma_{\xi_u}=d\xi_u + \vec{w}_{uv}\cdot d\vec{y}_v
\ee
where
\be
\vec\nabla_p \times \vec{w}_{uv} =\vec\nabla_p M_{uv}
\ee
$L_4 (=2\pi R_4)$ is the circumference of the $x_4$ circle of  D4 branes on $R^{1,3}\times S^1$ and   the mass
parameter $m_u$
\be
m_u = \frac{\epsilon_u}{L_4}
\ee
with $\sum^N_{u=1} \epsilon_u=1$
 is related to the Wilson line expectation value $\langle A_4\rangle$.
 The caloron system is related to the dynamics of  $N$ distinct monopoles sustained between
$T$-dual D3 branes, whose total magnetic charges
vanish.
$\vec{y}_u$ is the position 
of monopole (D-string)
connecting D3$_{u}$ and D3$_{u+1}$ (with D3$_N$ $=$D3$_0$) and the mass parameter
is then related to the mass of each distinct monopole. The positivity of $m_u$ implies that
we order the D3 brane locations along the $x_4$ direction monotonically.

Starting from this metric, we now derive the Calabi metric of $k=1$ U(N) instanton.
First we introduce the relative and the center-of-mass coordinates by
\be
\vec{x}_A = \vec{y}_{A A-1} \ \ (A=1,2,\cdots N-1)
\ee
and
\be
\vec{x}_N= \vec{x}_{\rm com}= \frac{\sum_u m_u \vec{y}_u}{\sum_u m_u}
\ee
and we shall denote this transformation by
\be
\vec{x}_u = U_{uv}\,  \vec{y}_v
\ee
Introducing
\be
\widetilde{M}= (U^T)^{-1} M U^{-1}
\ee
the metric becomes
\be
ds^2=\frac{4\pi^2}{g^2} L_4 \left[ \,
\widetilde{M}_{uv} d\vec{x}_u \cdot d\vec{x}_v +{\widetilde{M}}^{-1}_{uv} \sigma_{\varphi_u} \sigma_{\varphi_v}
\right]
\ee
where we introduce
\be
\varphi_u =\xi_v U^{-1}_{vu}\,, \ \ \ \
\sigma_{\varphi_u} = \sigma_{\xi_v} U^{-1}_{vu}
\ee
Then the derivative $\frac{\partial}{\partial \varphi_u}$ satisfies
$\frac{\partial}{\partial \varphi_u}=U_{uv}\, \frac{\partial}{\partial \xi_v }$, i.e.
\be
\frac{\partial}{\partial \varphi_A}= \frac{\partial}{\partial \xi_A }- \frac{\partial}{\partial \xi_{A-1} }
\ee
and
\be
\frac{\partial}{\partial \varphi_N}=  L_4 \sum_u m_u \frac{\partial}{\partial \xi_{u}}
\ee
where we have used the fact ${\sum_u m_u}=1/L_4$.
Note that   the charge $q_u\equiv  -2 i \frac{\partial}{\partial \xi_u }$ is integral quantized, i.e. $q_u \in \mathbb{Z}$ and
$Q_A\equiv -2 i \frac{\partial}{\partial \varphi_A}= q_A -q_{A-1} \in \mathbb{Z}$. Thus it is
clear that $\varphi_A$ is again ranged over $[0, 4\pi]$. One can check that
\be
 \sigma_{\varphi_u}=d\varphi_u + \vec{A}_{uv}\cdot d\vec{x}_v
\ee
where
\be
\vec\nabla_p \times \vec{A}_{uv} =\vec\nabla_p \widetilde{M}_{uv}
\ee
Let us introduce the relative mass $\mu_{AB}$ by
\be
\sum^N_{u=1} m_u \, d \vec{y}^2_u = \frac{d \vec{x}^2_{com}}{L_4}
+ \mu_{AB} d\vec{x}_A \cdot d\vec{x}_B
\ee
The metric then takes the form
\be
ds^2=\frac{4\pi^2 L_4}{g^2}  \left[
\frac{ d \vec{x}^2_{com}}{L_4} + {L_4} d \varphi^2_N +
\widetilde{M}_{AB} d\vec{x}_A \cdot d\vec{x}_B +{\widetilde{M}}^{-1}_{AB} \sigma_{\varphi_A} \sigma_{\varphi_B}
\right]
\ee
where
\be
\widetilde{M}_{AB} d\vec{x}_A \cdot d\vec{x}_B= \mu_{AB}d\vec{x}_A \cdot d\vec{x}_B
+\frac{d \vec{x}^2_{1}}{x_{1}}+\cdots +\frac{d \vec{x}^2_{N}}{x_{N}}
\ee

Now the Calabi limit is defined by taking the decompactification limit $L_4\rightarrow \infty$ with
the rescaling
\be
\vec{x}_A \rightarrow \frac{1}{L_4} \vec{x}_A
\ee
In this limit the metric becomes
\be
ds^2=\frac{4\pi^2 }{g^2}  \left[
 d {x}^i_{com}  d {x}^i_{com}+
C_{AB} d\vec{x}_A \cdot d\vec{x}_B +C^{-1}_{AB} \sigma_{\varphi_A} \sigma_{\varphi_B}
\right]
\ee
Here $x^i = (\vec{x}_{com},x^4_{com})$ where $x^4_{com}= L_4 \varphi_N$ is noncompact after taking the limit. The matrix $C_{AB}$ is as defined in (\ref{C}).

Since we are taking the decompactification limit, putting an electric charge to the monopole connecting D3$_N$ to D3$_1$ becomes
impossible. Hence in our interpretation of electric charges, the corresponding charge $q_N$ vanishes.

\section{Dyonic instanton BPS states in $\mathbb{R}^4$ }\label{Instanton}
The instanton 1/4 BPS equation localized to $\mb{R}^4$ reads
\bea
\Big[d-i_G+i(d^{\dag}- G )\Big]\Omega_\pm = 0
\eea
where
\bea
G &=& \frac{\bar{v}\bar{\rho}}{2} \bar{e}^3
\eea
using the notation of section \ref{internal}. We use the same letter for the one-form as for the corresponding dual vector field. We make the following Bose (even-form) ansatz
\bea
\Omega_{q\pm} &=& D_q \Lambda_{0\pm} + D_{q-1} \Lambda_{1\pm} + D_{q+1} \Lambda_{-1\pm}
\eea
where
\bea
\Lambda_{0\pm} &=& f_{\pm} (1\pm \bar{e}^0 \bar{e}^1 \bar{e}^2 \bar{e}^3) + g_{\pm} (\bar{e}^0 \bar{e}^3 \pm \bar{e}^1 \bar{e}^2)\cr
\Lambda_{1\pm} &=& c_{\pm} (\bar{e}^0\pm i \bar{e}^3)(\bar{e}^1+i\bar{e}^2)\cr
\Lambda_{-1\pm} &=& d_{\pm} (\bar{e}^0\mp i\bar{e}^3)(\bar{e}^1-i\bar{e}^2)
\eea
and $D_q$ denotes here the highest weight $D^j_{jq}$. Note that $j \ge q$.  The states with
\bea
D^j_{mq}= e^{-im \bar\psi-iq\bar\phi} d^j_{mq}(\bar\theta)
\eea
can be obtained by applying the lowering operator $\bar{L}_-$ using the
SU(2) rotational symmetry. Further using the rotational symmetry, the coefficient functions $ f_{\pm}$,  $ g_{\pm}$, $c_{\pm}$ and $d_{\pm}$ are only functions of $\bar{\rho}$.

By applying exterior derivative, we get
\bea
d\Lambda_{0\pm} &=& f'\bar{e}^0 \pm g' \bar{e}^0\bar{e}^1\bar{e}^2
- \delta_{down} \frac{4 g}{\bar\rho} \bar{e}^0 \bar{e}^1 \bar{e}^2 \cr
d\Lambda_{1\pm} &=& \pm c' i \bar{e}^0 \bar{e}^3 \bar{e}^+ -
 i \delta_{down} \frac{4c}{\bar\rho }  \bar{e}^0 \bar{e}^3 \bar{e}^+\cr
d\Lambda_{-1\pm} &=& \mp d' i  \bar{e}^0 \bar{e}^3 \bar{e}^- - i \delta_{up} \frac{4d}{\bar\rho }  \bar{e}^0 \bar{e}^3 \bar{e}^-
\eea
where we define $\delta_{down} = 1$ for lower sign and $\delta_{down}=0$ for upper sign, and $\delta_{up} = 1 - \delta_{down}$. We use
\bea
d D_q &=& i D_q \frac{2q}{\bar\rho } \bar{e}^3 + \frac{i\mu_q}{\bar\rho} D_{q-1} \bar{e}^+ + \frac{i\lambda_q}{\bar\rho} D_{q+1} \bar{e}^-
\eea
where
\bea
\mu_q &=& \sqrt{j(j+1)-q(q-1)}\cr
\lambda_q &=& \sqrt{j(j+1)-q(q+1)}
\eea
and
\bea
\bar{e}^3 \Lambda_{0\pm} &=& f \bar{e}^3 \pm g \bar{e}^1\bar{e}^2 \bar{e}^3\cr
\bar{e}^{\pm} \Lambda_{0\pm} &=& f \bar{e}^{\pm} + g \bar{e}^0 \bar{e}^3 \bar{e}^{\pm}\cr
\bar{e}^3 \Lambda_{1\pm} &=& -c \bar{e}^0 \bar{e}^3 \bar{e}^+\cr
\bar{e}^3 \Lambda_{-1\pm} &=& -d \bar{e}^0 \bar{e}^3 \bar{e}^-\cr
\bar{e}^- \Lambda_{1\pm} &=& -2ic (\bar{e}^0\pm i \bar{e}^3) \bar{e}^1\bar{e}^2\cr
\bar{e}^+ \Lambda_{-1\pm} &=& 2id (\bar{e}^0\mp i \bar{e}^3) \bar{e}^1 \bar{e}^2\cr
\bar{e}^+ \Lambda_{1\pm} &=& 0\cr
\bar{e}^- \Lambda_{-1\pm} &=& 0
\eea
and
\bea
* \bar{e}^3 &=& -\bar{e}^0 \bar{e}^1 \bar{e}^2\cr
* \bar{e}^0 &=& \bar{e}^1 \bar{e}^2 \bar{e}^3\cr
* \bar{e}^{\pm} &=& \mp i \bar{e}^0 \bar{e}^3 \bar{e}^{\pm}
\eea
and the fact that $** = -1$ on all these odd-dimensional forms.
We also note that
\bea
i_G \bar{e}^3 &=& \frac{\bar{v}\bar\rho }{2}\cr
i_G \bar{e}^0 &=& 0\cr
i_G \bar{e}^{\pm} &=& 0
\eea
We now find
\bea
d\Omega_{q\pm} &=& \bar{e}^3 iD_q\,  \frac{2q}{\bar\rho } f  \cr
&+& \bar{e}^0 D_q \, f'  \cr
&+& \bar{e}^1 \bar{e}^2 \bar{e}^3 i D_q \[\pm \frac{2q}{\bar\rho } g \pm \frac{\lambda_{q-1}}{\bar\rho}
2c \pm \frac{\mu_{q+1}}{\bar\rho}2d\]\cr
&+& \bar{e}^+ i D_{q-1} \, \frac{\mu_q}{\bar\rho} f  \cr
&+& \bar{e}^- i D_{q+1} \, \frac{\lambda_q}{\bar\rho}f  \cr
&+& \bar{e}^0 \bar{e}^3 \bar{e}^+ i D_{q-1} \[\frac{\mu_q}{\bar\rho} g - \frac{2(q-1)}{\bar\rho } c \pm c' -\delta_{down}
\frac{4c}{\bar\rho }\]\cr
&+& \bar{e}^0 \bar{e}^3 \bar{e}^- i D_{q+1} \[\frac{\lambda_q}{\bar\rho} g
- \frac{2(q+1)}{\bar\rho } d \mp d' -\delta_{up} \frac{4d}{\bar\rho }\]\cr
&+& \bar{e}^0 \bar{e}^1 \bar{e}^2 D_q \[\frac{\lambda_{q-1}}{\bar\rho} 2c -
\frac{\mu_{q+1}}{\bar\rho} 2d \pm  g' -\delta_{down} \frac{4g}{\bar\rho }\]
\eea
and
\bea
i * d\Omega &=& \bar{e}^3 i D_q \[\frac{\lambda_{q-1}}{\bar\rho} 2c - \frac{\mu_{q+1}}{\bar\rho} 2d \pm  g'
-\delta_{down} \frac{4g}{\bar\rho }\]\cr
&+& \bar{e}^0 D_q \[\pm {2q} g \pm \frac{\lambda_{q-1}}{\rho} 2c \pm \frac{\mu_{q+1}}{\rho}2d\]\cr
&+& \bar{e}^1 \bar{e}^2 \bar{e}^3 i D_q \, f'   \cr
&+& \bar{e}^+ i D_{q-1} \[\frac{\mu_q}{\bar\rho} g - \frac{2(q-1)}{\bar\rho } c \pm c' -\delta_{down} \frac{4c}{\bar\rho  }\]\cr
&+& (-)\bar{e}^- i D_{q+1} \[\frac{\lambda_q}{\bar\rho} g - \frac{2(q+1)}{\bar\rho } d \mp d'  -\delta_{up} \frac{4d}{\bar\rho  }\]\cr
&+& \bar{e}^0 \bar{e}^3 \bar{e}^+ i D_{q-1} \, \frac{\mu_q}{\bar\rho} f \cr
&+& (-) \bar{e}^0\bar{e}^3 \bar{e}^- i D_{q+1} \, \frac{\lambda_q}{\bar\rho}f \cr
&+& \bar{e}^0 \bar{e}^1 \bar{e}^2 D_q \, \frac{2q}{\bar\rho } f
\eea
We also have
\bea
iG\Omega &=& \frac{\bar{v}\bar\rho }{2} \[iD_q(f\bar{e}^3\pm g \bar{e}^1\bar{e}^2 \bar{e}^3) - i D_{q-1} c \bar{e}^0\bar{e}^3
\bar{e}^+ - i D_{q+1} d \bar{e}^0 \bar{e}^3 e^-\]
\nonumber
\eea
and
\bea
i_G \Omega &=& \frac{v\rho K}{2} \[D_q(\mp f e^0 e^1 e^2 - g e^0) \pm i D_{q-1} c e^+ \mp i D_{q+1} d e^-\]
\eea
The BPS equation is
\bea
d\Omega - i_G \Omega &=& \pm i *d\Omega +iG\Omega
\eea
that we write as
\bea
d\Omega \mp i *d\Omega &=& iG\Omega + i_G \Omega
\eea
The LHS now becomes
\bea
&& (\bar{e}^3 i \mp  \bar{e}^0 \bar{e}^1 \bar{e}^2)D_q \[- g'-4 \delta_{down} \frac{g }{\bar\rho}+\frac{2q}{\bar\rho } f \mp  \frac{\lambda_{q-1}}{\bar\rho} 2c \pm \frac{\mu_{q+1}}{\bar\rho} 2d \] \cr
&+& (\bar{e}^0-\bar{e}^1 \bar{e}^2 \bar{e}^3 i)D_q \[f' - \frac{2q}{\bar\rho } g -
 \frac{\lambda_{q-1}}{\bar\rho} 2c -\frac{\mu_{q+1}}{\bar\rho}2d\]\cr
&+& (\bar{e}^+ i \mp \bar{e}^0 \bar{e}^3 \bar{e}^+ i) D_{q-1} \[- c'-4 \delta_{down} \frac{c}{\bar\rho }+\frac{\mu_q}{\bar\rho}
(f\mp g) \pm \frac{2(q-1)}{\bar\rho } c \] \cr
&+& (\bar{e}^- i \pm \bar{e}^0 \bar{e}^3 \bar{e}^- i)D_{q+1} \[- d' -4 \delta_{up} \frac{d}{\bar\rho }
+\frac{\lambda_q}{\bar\rho}(f \pm g) \mp \frac{2(q+1)}{\bar\rho } d \]
\eea
The RHS is the sum of
\bea
iG\Omega &=& \frac{\bar{v}\bar\rho }{2} \[iD_q(f \bar{e}^3\pm g \bar{e}^1\bar{e}^2 \bar{e}^3)
 - i D_{q-1} c\, \bar{e}^0 \bar{e}^3 \bar{e}^+ - i D_{q+1} d \, \bar{e}^0 \bar{e}^3 \bar{e}^-\]
\nonumber
\eea
and
\bea
i_G \Omega &=& \frac{\bar{v}\bar\rho }{2} \[D_q(\mp f \bar{e}^0 \bar{e}^1 \bar{e}^2 -
g \bar{e}^0) \pm i D_{q-1} c \, \bar{e}^+ \mp i D_{q+1} d\, \bar{e}^-\]
\eea
Thus RHS is
\bea
\frac{\bar{v}\bar\rho }{2}&\times & \Bigg[(\bar{e}^3 i \mp \bar{e}^0 \bar{e}^1 \bar{e}^2)D_q \, f  \cr
&+& (\bar{e}^0 \mp \bar{e}^1\bar{e}^2\bar{e}^3 i)D_q (-)g \cr
&+& (\bar{e}^+ i \mp \bar{e}^0 \bar{e}^3 \bar{e}^+ i )D_{q-1}  (\pm) c \cr
&+& (\bar{e}^- i \mp \bar{e}^0\bar{e}^3 \bar{e}^- i )D_{q+1}  (\mp) d\Bigg]
\eea
Subtracting RHS - LHS, we have
\bea
&& (\bar{e}^3 i \mp  \bar{e}^0 \bar{e}^1 \bar{e}^2)D_q \[- g'
+\frac{2q}{\bar\rho } f \mp  \frac{\lambda_{q-1}}{\bar\rho} 2c \pm \frac{\mu_{q+1}}{\rho} 2d - \frac{\bar{v}\bar\rho }{2}f - \delta_{down}\frac{4}{\bar\rho}g\] \cr
&+& (\bar{e}^0-\bar{e}^1 \bar{e}^2 \bar{e}^3 i)
D_q \[f'  - {2q}\, g - \frac{\lambda_{q-1}}{\bar\rho} 2c - \frac{\mu_{q+1}}{\bar\rho}2d + \frac{\bar{v}\bar\rho }{2}g\]\cr
&+& (\bar{e}^+ i \mp \bar{e}^0 \bar{e}^3 \bar{e}^+ i) D_{q-1} \[- c'
+\frac{\mu_q}{\bar\rho} (f\mp g) \pm \frac{2(q-1)}{\bar\rho } c \mp \frac{\bar{v}\bar\rho }{2}c -
\delta_{down}\frac{4c}{\bar\rho }\] \cr
&+& (\bar{e}^- i \pm  \bar{e}^0\bar{e}^3 \bar{e}^- i)D_{q+1}
\[- d'+\frac{\lambda_q}{\bar\rho}(f \pm g) \mp \frac{2(q+1)}{\bar\rho } d \pm \frac{\bar{v}\bar\rho }{2}d
- \delta_{up} \frac{4d}{\bar\rho }\] \nonumber
\eea
Thus for upper sign (SD case) we have the BPS equations
\bea
- g' +\frac{2q}{\bar\rho } f - \frac{\lambda_{q-1}}{\bar\rho } 2c - \frac{\mu_{q+1}}{\bar\rho } 2d - \frac{\bar{v}\bar\rho}{2}f
&=& 0 \cr
f' - \frac{2q}{\bar\rho } g - \frac{\lambda_{q-1}}{\bar\rho } 2c +\frac{\mu_{q+1}}{\bar\rho }2d + \frac{\bar{v}\bar\rho}{2}g &=& 0\cr
- c'+\frac{\mu_q}{\bar\rho } (f - g) + \frac{2(q-1)}{\bar\rho } c - \frac{\bar{v}\bar\rho}{2}c  &=& 0 \cr
- d'+\frac{\lambda_q}{\bar\rho}(f + g) - \frac{2(q+1)}{\bar\rho} d + \frac{\bar{v}\bar\rho}{2}d -  \frac{4}{\bar\rho}d &=& 0
\eea
and for lower sign (ASD case) we have the BPS equations
\bea
- g' +\frac{2q}{\bar\rho} f +  \frac{\lambda_{q-1}}{\bar\rho} 2c - \frac{\mu_{q+1}}{\bar\rho} 2d
- \frac{\bar{v}\bar\rho }{2}f -\frac{4g}{\bar\rho} &=& 0 \cr
f'  - \frac{2q}{\bar\rho} g - \frac{\lambda_{q-1}}{\bar\rho} 2c - \frac{\mu_{q+1}}{\bar\rho}2d + \frac{\bar{v}\bar\rho}{2}g &=& 0\cr
- c'+\frac{\mu_q}{\bar\rho} (f + g) - \frac{2(q-1)}{\bar\rho} c + \frac{\bar{v}\bar\rho}{2}c - \frac{4}{\bar\rho}c &=& 0 \cr
- d'+\frac{\lambda_q}{\bar\rho}(f - g) + \frac{2(q+1)}{\bar\rho} d - \frac{\bar{v}\bar\rho}{2}d &=& 0
\eea
For the upper sign, we define $h = f+g$ and $s=f-g$. Then the equations
 become
\bea
s'- \frac{\bar{v}\bar\rho }{2}s +\frac{2q}{\bar\rho} s - \frac{\lambda_{q-1}}{\bar\rho} 4c &=& 0 \cr
c'+ \frac{\bar{v}\bar\rho}{2}c - \frac{2(q-1)}{\bar\rho} c  -\frac{\mu_{q}}{\bar\rho}s  &=& 0 \cr
h'+ \frac{\bar{v}\bar\rho }{2}h -\frac{2q}{\bar\rho} h - \frac{\mu_{q+1}}{\bar\rho} 4d &=& 0 \cr
d'- \frac{\bar{v}\bar\rho}{2}d + \frac{2(q+1)}{\bar\rho} d  - \frac{\lambda_q}{\bar\rho}h  &=& 0
\eea


First consider the case of  $j > q$. One finds that $\lambda_{q-1}=\mu_q$ and $\lambda_q=\mu_{q+1}$ are
all nonvanishing.
Requiring the normalizability of the wave function, one can show that $s=c=h=d=0$. Hence such states
do not exist.
The brief discussion of the proof is as follows. Let us consider the first two coupled
equations. Eliminating $s$, one finds
\bea
\left[ \Big(\bar\rho \frac{d}{d\bar\rho}\Big)^2 +2\bar\rho \frac{d}{d\bar\rho}
-\frac{\bar{v}^2 \bar\rho^4}{4}+2q \bar{v} \bar\rho^2 -2q(q-1)
\right] c= 4 \mu_q^2 c
\eea
One can rearrange this equation to the following form
\bea
\left[- \frac{1}{\bar\rho^3} \frac{d}{d\bar\rho}\bar\rho^3 \frac{d}{d\bar\rho}
+\Big(\frac{\bar{v} \bar\rho}{2}-\frac{2q}{\bar\rho}\Big)^2
\right] c= -4 \frac{(\mu_q^2 -q)}{\bar\rho^2} c
\label{normal}
\eea
Note that $\mu^2_q -q > 0$ since we are interested in coupled case requiring
$j \ge q$ when $q \ge \frac{1}{2}$ and $j \ge 1$ if $q=0$. Performing an integration with respect to $\bar\rho$ after multiplying (\ref{normal}) by
 $\bar\rho^3 c^*$
(i.e. $\int^\infty_0 d\bar\rho \bar\rho^3 c^* $ (\ref{normal})) , one finds that the LHS is positive definite
while the RHS is negative definite unless $c=0$. Therefore we conclude that $s=c=0$ once the two equations are coupled with each
other.
A similar argument goes through for the latter two equations leading to $h=d=0$.

For $j=q$, $\lambda_{q-1}$ and $\mu_q$ are nonvanishing while  $\lambda_q=\mu_{q+1}=0$. Then again requiring the
normalizability, one finds $s=c=d=0$ and
\bea
f=g=  \bar\rho^{2q} e^{-\frac{\bar{v} \bar\rho^2}{4}}
\eea
which leads to the $j=q$ multiplet. For $j=q-1$, $h=s=d=0$ by definition. One has
\bea
c= \bar\rho^{2(q-1)} e^{-\frac{\bar{v} \bar\rho^2}{4}}
\eea

Similarly one can show that the ASD sector does not have any solution based on the fact
that four are all coupled among themselves.

For the case with negative $\bar{v}$, $q$ has to be non-positive definite. One finds the following solutions
by the same argument as the positive  $\bar{v}$ case in the above.
For $j=|q|$, one finds
\bea
f=-g=  \bar\rho^{2|q|} e^{-\frac{|\bar{v}| \bar\rho^2}{4}}
\eea
with $c=d=0$.
For $j=|q+1|$ with $q \le -1$, one has
\bea
d= \bar\rho^{2(|q+1|)} e^{-\frac{|\bar{v}| \bar\rho^2}{4}}
\eea
with $f=g=c=0$.

\section{Eguchi-Hanson metric}\label{SEguchi}
For $N=2$ we can directly solve the ADHM constraints as
\ben
(w_{\dot{1}u} \quad w_{\dot{2}u}) &=& g \(\begin{array}{cc}
\sqrt{\rho^2 + 2\zeta} & 0\\
0 & \sqrt{\rho^2 - 2\zeta}
\end{array}\)\label{Eguchi}
\een
where
\bea
g &=& \(\begin{array}{cc}
u_1  & -u^2 \\
u_2 & u^1
\end{array}\)
\eea
and
\ben
u^1 &=& \cos \frac{\theta}{2} e^{i(\psi + \phi)/2}\cr
u^2 &=& \sin \frac{\theta}{2} e^{i(\psi -\phi)/2}\label{param}
\een
We can in addition make a $U(1)$ gauge transformation
\bea
{w}_{\dot\alpha u} &\rightarrow & e^{-i\xi} {w}_{\dot\alpha u}
\eea
of our solution. After this transformation, the metric (\ref{flat1}) induces the metric
\bea
ds^2 &=& 2\rho^2 \(d\xi + \frac{\alpha}{2\rho^2}\)^2 + d w_{\dot\alpha u} d\bar w^{\dot \alpha u} - \frac{\alpha^2}{2\rho^2}
\eea
where
\bea
\alpha &=& -i w_{\dot \alpha u} d \bar w^{\dot\alpha u} + i dw_{\dot\alpha u} \bar w^{\dot \alpha u}
\eea
Here $\xi$ is a cyclic coordinate, in the sense that the metric does not depend on $\xi$ (but only on $d\xi$). The conjugate momentum
\bea
p_{\xi} &=& d\xi + \frac{\alpha}{2\rho^2}
\eea
to $\xi$ is a conserved quantity. Modding out by $U(1)$ gauge symmetry amounts to putting $p_{\xi} = 0$. An alternative approach is to define a covariant derivative
\bea
Dw_{\dot\alpha u} &=& dw_{\dot \alpha u} - i A w_{\dot\alpha u}
\eea
define the moduli space metric as
\bea
ds^2 &=& Dw_{\dot\alpha u} D\bar w^{\dot\alpha u}
\eea
and use the Gauss law constraint, which amounts to extremizing this metric with respect to $A$. Either way the moduli space metric becomes
\bea
ds^2 &=& 2 \(\frac{1}{1 - \frac{4\zeta^2}{\rho^4}} d\rho^2 + \rho^2 du_{\alpha} du^{\alpha} - \frac{4\zeta^2}{\rho^2} \(iu^{\alpha} du_{\alpha}\)^2\)
\eea
From (\ref{param}) we get
\ben
du_{\alpha} du^{\alpha} &=& \frac{1}{4} \(d\theta^2 + \sin^2 \theta d\phi^2 + (d\psi + \cos \theta d\phi)^2\)\cr
iu^{\alpha} du_{\alpha} &=& -\frac{1}{2}\(d\psi + \cos \theta d\phi\)
\een
and thus
\bea
ds^2 &=& 2\(\frac{d\rho^2}{1 - \frac{4\zeta^2}{\rho^4}} + \frac{\rho^2}{4} \(\sigma_1^2 + \sigma_2^2 + \(1-\frac{4\zeta^2}{\rho^4}\)\sigma_3^2\)\)
\eea
where
\bea
\sigma_1^2 + \sigma_2^2 &=& d\theta^2 + \sin^2 \theta d\phi^2 \cr
\sigma_3 &=& d\psi + \cos \theta d\phi
\eea
This is the Eguchi-Hanson metric.

\subsection{Coordinate map from Eguchi-Hanson to Calabi metric}
We have seen two different ways to obtain the moduli space metric from ADHM constraints by factoring out the $U(1)$ gauge symmetry. For $N=2$ the two methods must give the same moduli space, which means that the Calabi metric must be equivalent to the Eguchi-Hanson metric by means of a coordinate transformation. To find the coordinate transformation we will now more carefully compare the two methods we used to obtain these two metrics.

In the Calabi case and for $N=2$ we eliminate $\x_2$ by expressing it in terms of the Calabi space coordinates
\bea
\x_{1} &=& \frac{1}{4} w_{1 \dot\alpha} \vec{\sigma}^{\dot \alpha}{}_{\dot\beta} \bar w^{\dot\beta}_1
\eea
From (\ref{Eguchi}) we have
\bea
w_{1\dot\alpha} &=& (u_1 \rho_+, -u^2 \rho_-)
\eea
where we define $\rho_{\pm} = \sqrt{\rho^2 \pm 2\zeta}$. Using the parametrization (\ref{param}), we get
\bea
x^1_1 &=& \frac{1}{4}\(-\rho_+ \rho_- \sin \theta \cos \psi\)\cr
x^2_1 &=& \frac{1}{4}\(\rho_+ \rho_- \sin \theta \sin \psi\)\cr
x^3_1 &=& \frac{1}{4}\(\rho^2 \cos \theta + 2\zeta\)
\eea
and we may recall that $x_2^1 = - x_1^1$, $x_2^2 = -x_1^2$ and $x_2^3 = \zeta - x_1^3$. It is convenient to introduce the notation $z_+ = x^3_1$ and $z_- = x^3_2$. Then
\bea
z_{\pm} &=& \frac{1}{4}\(\pm \rho^2 \cos \theta + 2\zeta\)\cr
r_{\pm} &=& \sqrt{x_1^2 + x_1^2 + z_{\pm}^2}
\eea
Quite remarkably we find the square root of a perfect square
\bea
r_{\pm} &=& {\lambda}\(\rho^2 \pm 2\zeta \cos \theta\)
\eea
From (\ref{C}) we then get
\bea
C &=& \frac{2\rho^2}{\lambda\(\rho^4 - (2\zeta)^2 \cos^2 \theta\)}
\eea
The fiber coordinate and gauge potential in Calabi coordinates is given by
\bea
\varphi &=& \psi_1 - \psi_2\cr
A &=& A_1 - A_2
\eea
respectively, where
\bea
A_u &=& \frac{1}{r_u \(r_u - z_u\)}\(x_u dy_u - y_u dx_u\)
\eea
In Eguchi-Hanson coordinates the Calabi space gauge potential becomes
\bea
A &=& \frac{2(\rho^4 - (2\zeta)^2) \cos \theta}{\rho^4 - (2\zeta)^2 \cos^2 \theta} d\psi
\eea
We shall identify
\bea
\varphi &=& 2\phi
\eea
in order to match Calabi metric with Eguchi-Hanson metric. Thus the fiber direction which is parametrized by $\psi$ in the Eguchi-Hanson metric shall not to be confused with the fiber over $\mb{R}^3$ of the Calabi metric which is parameterized by $\varphi$.

\subsection{Properties of Calabi space}
The Eguchi-Hanson metric gives us the vierbein
\bea
e^1 &=& \frac{\rho}{2} \sigma^1\cr
e^2 &=& \frac{\rho}{2} \sigma^2\cr
e^3 &=& \frac{\rho}{2} K \sigma^3\cr
e^4 &=& \frac{d\rho}{K}
\eea
up to a local $SO(4)$ rotation. We define the spin connection $\omega^{ab}$ by
\bea
de^a + \omega^{ab} e^b &=& 0
\eea
Defining $\epsilon^{1234} = 1$ we have \cite{Eguchi:1978gw}
\bea
\omega^{ab} &=& \frac{1}{2} \epsilon^{abcd} \omega^{cd}
\eea
The three Kahler forms are now given by
\bea
I^I &=& \frac{1}{2} \eta^{I-}_{ab} e^a \wedge e^b
\eea
We can see that these are closed by the following argument. We first obtain
\bea
dI^I &=& \eta^{I-}_{ab} \omega^{ac} e^c \wedge e^b
\eea
We then expand the selfdual spin connection as $\omega^{ac} = \eta^{J+}_{ac} \omega^J$ and we find that
\bea
dI^I = [\eta^{I-},\eta^{J+}]_{ab} e^a \wedge e^b = 0
\eea

To obtain Kahler forms on Calabi space we start with flat space $\mb{H}^N = \mb{C}^{2N}$ with metric (\ref{metric1}). Three Kahler forms are given by
\bea
I^I &=& \sum_{u=1}^N \eta^{I+}_{ab} e^a_u \wedge e^b_u
\eea
where the vielbein is given by
\bea
e^I_u &=& C_u^{\frac{1}{2}} dx^I_u\cr
e^4_u &=& C_u^{-\frac{1}{2}} \sigma_{\psi_u}
\eea
By using
\bea
d\sigma_{\psi_u} &=& \frac{1}{2x_u^3} \epsilon^{IJK} x_u^I dx_u^J \wedge dx_u^K
\eea
we may check that
\bea
dI^I &=& 0
\eea
To derive the Kahler forms on Calabi space we eliminate $\vec{x}_N$ using the ADHM constraint (\ref{ADHMC}) and we define $\sigma_{\varphi_A}$ as in (\ref{varphi}). We then get
\bea
I^I &=& dx_A^I \wedge \sigma_A + \frac{1}{2} \epsilon^{IJK} C_{AB} dx^J_A \wedge dx^K_B
\eea

\end{document}